\documentclass[prd,superscriptaddress,twocolumn,nofootinbib,showpacs,preprintnumbers,floatfix]{revtex4}

\usepackage{mathrsfs}
\usepackage{amsfonts,amssymb,amsmath}
\usepackage{graphicx,epsfig}
\usepackage{multirow}

\begin{document}

\title{The ${\cal F}$-Landscape: Dynamically Determining the Multiverse}

\author{Tianjun Li}

\affiliation{Key Laboratory of Frontiers in Theoretical Physics, Institute of Theoretical Physics,
Chinese Academy of Sciences, Beijing 100190, P. R. China }

\affiliation{George P. and Cynthia W. Mitchell Institute for Fundamental Physics and Astronomy,
Texas A$\&$M University, College Station, TX 77843, USA }

\author{James A. Maxin}

\affiliation{George P. and Cynthia W. Mitchell Institute for Fundamental Physics and Astronomy,
Texas A$\&$M University, College Station, TX 77843, USA }

\author{Dimitri V. Nanopoulos}

\affiliation{George P. and Cynthia W. Mitchell Institute for Fundamental Physics and Astronomy,
Texas A$\&$M University, College Station, TX 77843, USA }

\affiliation{Astroparticle Physics Group, Houston Advanced Research Center (HARC),
Mitchell Campus, Woodlands, TX 77381, USA}

\affiliation{Academy of Athens, Division of Natural Sciences,
28 Panepistimiou Avenue, Athens 10679, Greece }

\author{Joel W. Walker}

\affiliation{Department of Physics, Sam Houston State University,
Huntsville, TX 77341, USA }


\begin{abstract}

We evolve our \textit{Multiverse Blueprints} to characterize our local neighborhood of the String Landscape and the Multiverse of
plausible string, M- and F-theory vacua. Building upon the tripodal foundations of i) the Flipped $SU(5)$ Grand Unified Theory (GUT),
ii) extra TeV-Scale vector-like multiplets derived out of F-theory, and iii) the dynamics of No-Scale Supergravity, together dubbed
No-Scale ${\cal F}$-$SU(5)$, we demonstrate the existence of a continuous family of solutions which might adeptly describe
the dynamics of distinctive universes.  This Multiverse landscape of ${\cal F}$-$SU(5)$ solutions, which we shall refer to as the
${\cal F}$-Landscape, accommodates a subset of universes compatible with the presently known experimental uncertainties of our own
universe. We show that by secondarily minimizing the minimum of the scalar Higgs potential of each solution within the ${\cal F}$-Landscape,
a continuous hypervolume of distinct \textit{minimum minimorum} can be engineered which comprise a regional dominion of universes,
with our own universe cast as the bellwether.  We conjecture that an experimental signal at the LHC of the No-Scale ${\cal F}$-$SU(5)$
framework's applicability to our own universe might sensibly be extrapolated as corroborating evidence for the role of string, M- and
F-theory as a master theory of the Multiverse, with No-Scale supergravity as a crucial and pervasive reinforcing structure.

\end{abstract}


\preprint{ACT-18-11, MIFPA-11-49}

\maketitle


\section{Introduction}

Contemporary times have witnessed a revolution in string phenomenology, the culmination of decades of enterprise
toward the comprehension of a fundamental high energy theory capable of describing the evolution of our observable universe.
An unwavering theme that has emerged from this century of innovation is nature's persistent rejection of an intransigent
conception of the macrocosm, of which we are just a simple element.  Nature's truths have been revealed in pieces and in
paradoxes, and have stymied every effort to claim mastery over her mysteries.  Whether it be relativistic space and time,
quantum entanglement, or black hole event horizons, we have become acclimated to radical revisions in our sense of reality,
recognizing that the course of time may force all to acquiesce to axioms initially seeming exotic and fantastic,
if they be first synthesized upon rigorous physical maxims.

Progress in the understanding of consistent, meta-stable vacua of string, M- or (predominantly) F-theory flux compactifications
has inspired dramatic challenges to the perspective of our prominence in the cosmos.  Case in point, it has been postulated that
a vast landscape of an astonishing $10^{500}$~\cite{Denef:2004dm,Denef:2007pq} vacua can manifest plausible phenomenology in general.
This suggestion implores inquiry as to why our peculiar vacuum transpired out of the landscape.  One prevalent philosophy contends that
any physically existent universe, whether latent or mature, should correspond to an extremization of probability density in the primordial
quantum froth.  Known as the \textit{Anthropic Principle}, this idea implies that our universe, due to its natural existence and presumed
singularity, occupies a statistical zenith.  Consequently though, this doctrine becomes incurably burdened with fine-tuning complications
of the physical properties of our universe. Motivated by the string landscape and other cosmological scenarios, the speculation of a
Multiverse germinated as a strategy for overcoming those obstacles endemic to fine-tuning.

In our contemporary \textit{Multiverse Blueprints}~\cite{Li:2011dw} we advanced an alternate perspective of our cosmological origins.
We suggested that a mere non-zero probability for a universe featuring our measured physical parameters is the necessary and sufficient
condition.  An observer may inhabit a universe bearing simply a probability of existence which is greater than zero, and not inevitably
that which is most probable.  Moreover, we argued for the significance of No-Scale Supergravity as a universal foundation allowing
for the spontaneous quantum emergence of a cosmologically flat universe.  Experimental validation of a No-Scale ${\cal F}$-$SU(5)$ structure
for our own universe at the LHC could thus reinforce the role of string, M- and F-theory as a master theory of the Multiverse, with No-Scale
supergravity providing an essential model building infrastructure.

\begin{figure*}[htp]
	\centering
	\includegraphics[width=1.00\textwidth]{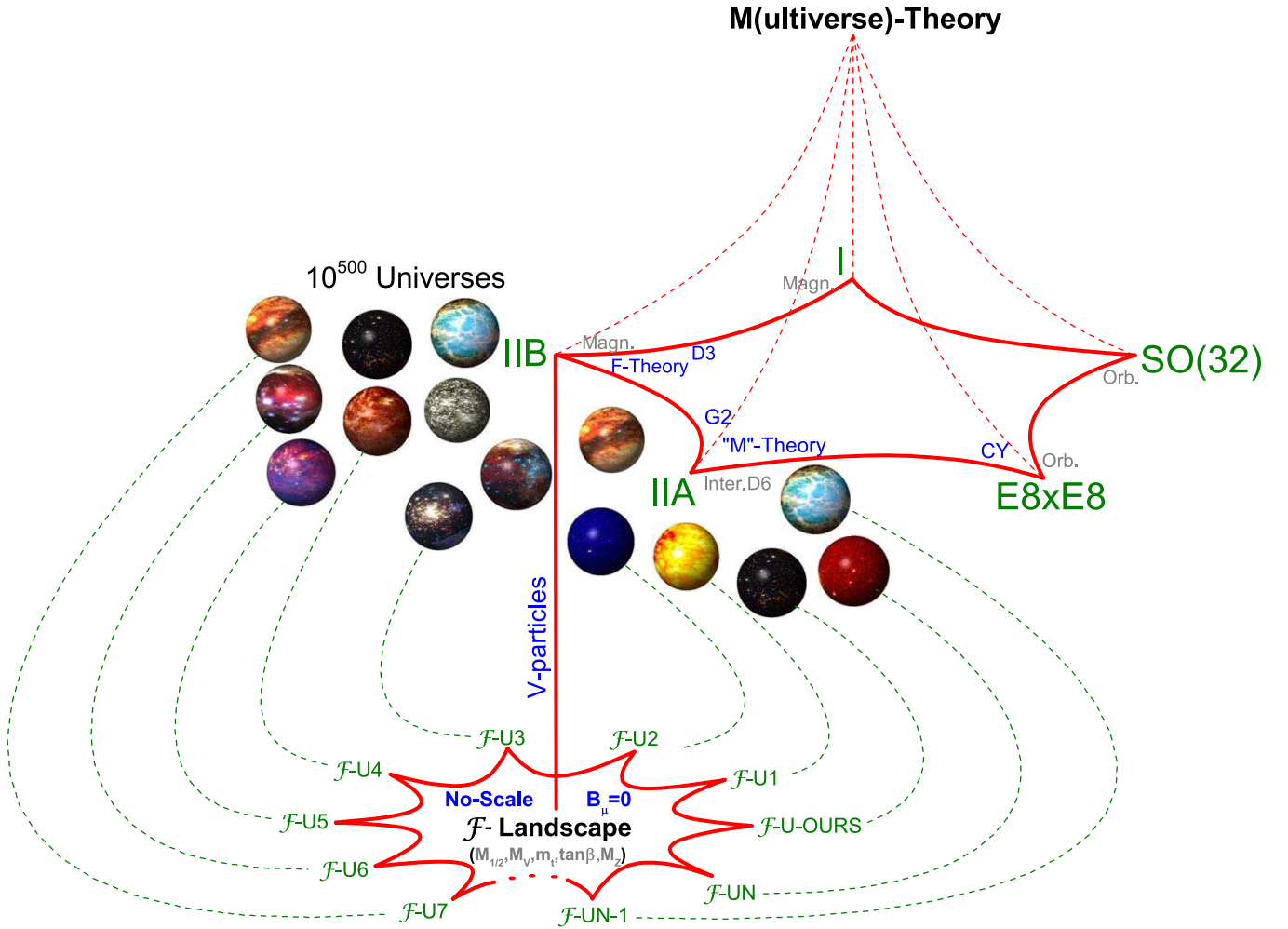}
	\caption{Contemporary perspective on the String Landscape and M-Theory, where we build the M(ultiverse)-Theory with the ${\cal F}$-Landscape derived out of the tripodal foundation in Fig.~\ref{fig:MVII_MultiversePyramid}.}
	\label{fig:MVII_FLandscape_Stars}
\end{figure*}

\begin{figure*}[htp]
	\centering
	\includegraphics[width=1.00\textwidth]{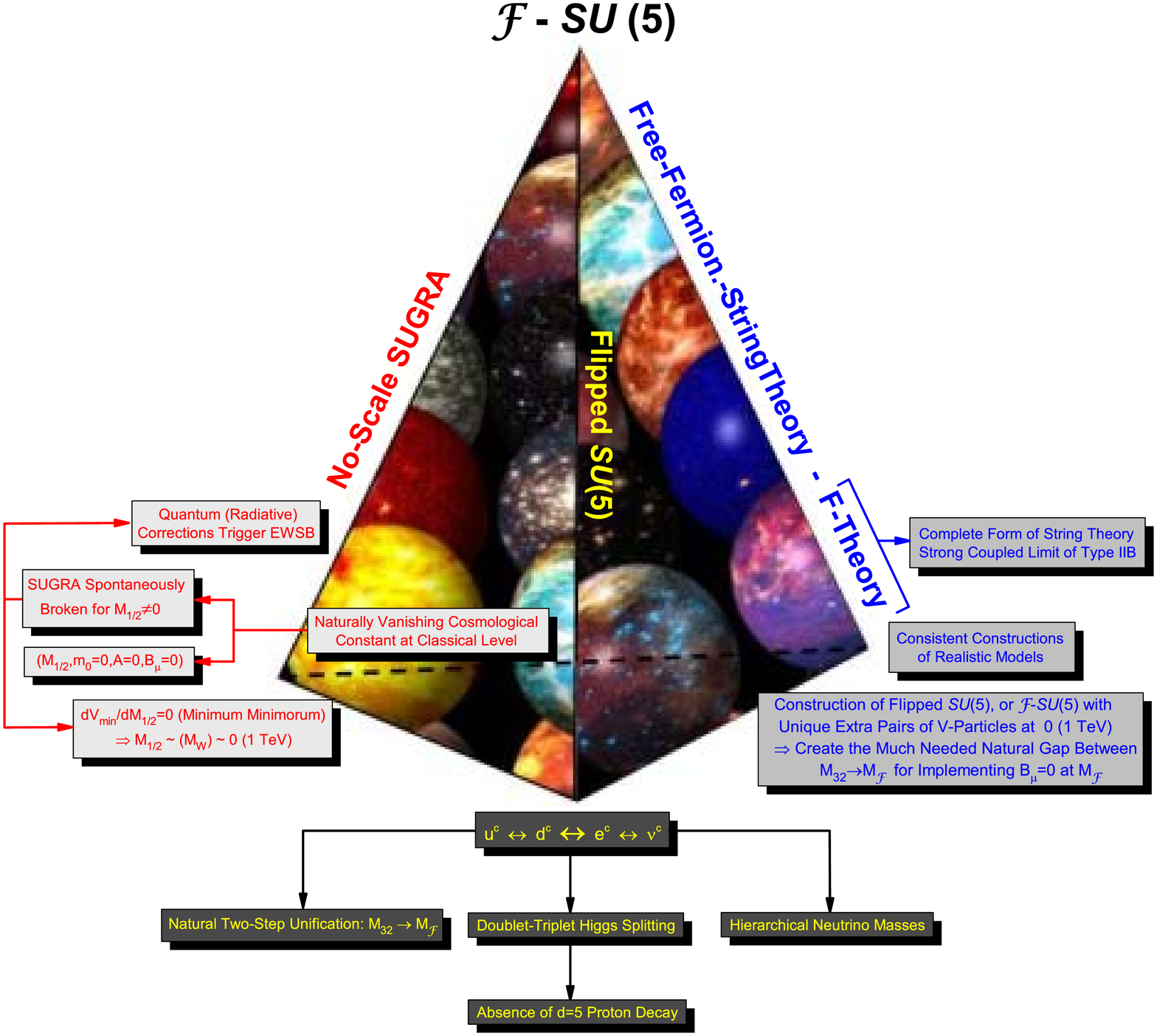}
	\caption{Tripodal foundation of ${\cal F}$-$SU(5)$, built upon the Flipped $SU(5)$ Grand Unified Theory (GUT), extra TeV-Scale vector-like multiplets derived out of F-theory, and the dynamics of No-Scale Supergravity.}
	\label{fig:MVII_MultiversePyramid}
\end{figure*}

We now undertake a first task of engineering in association with our Blueprints~\cite{Li:2011dw}, considering the possibility of Multiverse model building,
or \textit{universe building}.  We employ a precision numerical analysis to derive and subsequently classify the features of the No-Scale ${\cal F}$-$SU(5)$
Multiverse, within some local neighborhood of our own universe's phenomenology.  By secondarily minimizing each model's scalar Higgs potential minima, under
application of the dynamic Super No-Scale condition~\cite{Li:2010uu,Li:2011dw,Li:2011xu}, only legitimate electroweak symmetry breaking (EWSB) vacua are
viable elements of the solution space. The dynamically selected EWSB vacuum at this point of secondary minimization, which is in correspondence with the stabilization
of a string-theoretic modulus, will be identified as the \textit{minimum minimorum} (MM).  Thereupon, \textit{all} MM realize our minimal specifications for a greater
than zero probability of emerging from the landscape.  Hence, we conclude that a contiguous hyperspace of MM in No-Scale ${\cal F}$-$SU(5)$ may fulfill the intended
goal of constructing the set of locally adjacent Multiverse constituents, endogenous to the plausible solution set of M- and F-Theory flux compactifications. We stress
that application of the dynamic MM vacuum selection criterion elevates the conceptual Multiverse design presented here above a mere scan of the parameter space.
We suggest that the resulting construction might rather be regarded to represent a local dominion of independent universes.

Our paper is organized as follows. First, we shall discuss No-Scale ${\cal F}$-$SU(5)$ in M- and F-Theory flux compactifications,
presenting our ${\cal F}$-$SU(5)$ M(ultiverse)-Theory. Next, we engage in a brief review of ${\cal F}$-$SU(5)$, the Super No-Scale condition,
and our secondary minimization procedure. In the latter half of our work, we shall demonstrate the minimization of discrete elements within the model space,
and extrapolate the results to construct a hyperspace of MM, interpreting the solution space in terms of our local community of universes within the Multiverse.

\section{The ${\cal F}$-$SU(5)$ M(ultiverse)-Theory}

The Standard Model has been confirmed as a correct effective field
theory valid up to about 100~GeV. Nonetheless, problems exist, such as the gauge hierarchy problem,
charge quantization, and an excessive number of parameters, etc. Moreover, the Standard Model excludes gravity. An elegant solution to the gauge hierarchy
problem is supersymmetry. In particular, gauge coupling
unification can be realized in the supersymmetric SM (SSM), which
strongly implies the Grand Unified Theories (GUTs). In the
GUTs, not only can we explain the charge quantization, but
also reduce the Standard Model parameters due to unification.
Therefore, the interesting question is whether there exists a fundamental
quantum theory or a final theory that can unify the SSM/GUTS 
and general relativity?

The most promising candidate for such a theory is superstring theory.
Superstring theory is anomaly free only in ten dimensions, hence
the extra six space dimensions must be compactified.
As portrayed in Fig.~\ref{fig:MVII_FLandscape_Stars}, there are five consistent ten-dimensional superstring theories:
 heterotic $E_8 \times E_8$, heterotic $SO(32)$, Type I $SO(32)$, Type IIA, 
and Type IIB. Though, this leaves open the question of final unification.
Interestingly, Witten pointed out that this distinction is an artifact of
perturbation theory, and non-perturbatively these five superstring theories
are unified into an eleven-dimensional M-theory~\cite{Witten:1995ex}.
In other words, the five superstring theories are the different perturbative
limits of M-theory. Moreover,  the twelve-dimensional 
F-theory can be considered as the strongly coupled formulation of the Type IIB
string theory with a varying axion-dilaton field~\cite{Vafa:1996xn}, as shown in Fig.~\ref{fig:MVII_FLandscape_Stars}. 

The goal of string phenomenology is to construct the realistic
string vacua, where the SSM/GUTs can be realized and the moduli
fields can be stabilized. Such constructions will give us a bridge
between the string theory and the low energy realistic particle
physics, such that we may test the string models at the Large
Hadron Collider (LHC). Initially, string phenomenology
was studied mainly in the weakly coupled heterotic string theory.
On the other hand, we illustrate in Fig.~\ref{fig:MVII_FLandscape_Stars} that in addition to its perturbative
heterotic string theory corner, M-Theory unification possesses the other corners such as perturbative Type I, Type IIA
and Type IIB superstring theory, which should provide new potentially phenomenologically
interesting four-dimensional string models, related to the heterotic models via 
a web of string
dualities. Most notably, with the advent of D-branes~\cite{Polchinski:1995mt}, 
we can construct the 
phenomenologically interesting string models in Type I, Type
IIA and Type IIB string theories.
Recall that there are five kinds of string models which have 
been studied extensively: (1)  Heterotic $E_8\times E_8$ string model building. The
supersymmetric SM and GUTs can be constructed via
the orbifold 
compactifications~\cite{Buchmuller:2005jr, Lebedev:2006kn, Kim:2006hw} 
and the Calabi-Yau manifold 
compactifications~\cite{Braun:2005ux, Bouchard:2005ag};
(2) Free fermionic string model building. Realistic models
with clean particle spectra can only be constructed at 
the Kac-Moody level one~\cite{Antoniadis:1987tv, 
Antoniadis:1988tt, Antoniadis:1989zy,
 Faraggi:1989ka,
Antoniadis:1990hb, Lopez:1992kg, Cleaver:2001ab}. Note that the Higgs
fields in the adjoint representation or higher can not be
generated at the Kac-Moody level one, so only three kinds
of models can be constructed: the Standard-like models,
Pati-Salam models, and flipped $SU(5)$ 
models~\cite{Antoniadis:1987tv, 
Antoniadis:1988tt, Antoniadis:1989zy,
 Faraggi:1989ka,
Antoniadis:1990hb, Lopez:1992kg, Cleaver:2001ab}.
(3) D-brane model building from Type I, Type IIA,
and Type IIB theories. There are two major kinds of such
models: (i) Intersecting D-brane models or magnetized
D-brane models~\cite{Berkooz:1996km,
Ibanez:2001nd, Blumenhagen:2001te, Cvetic:2001tj,
Cvetic:2001nr, Cvetic:2002pj, Cvetic:2004ui, Cvetic:2004nk,
 Chen:2005aba, Chen:2005mm, Chen:2005mj, Blumenhagen:2005mu};
(ii) Orientifolds of Gepner 
models~\cite{Dijkstra:2004ym, Dijkstra:2004cc}. 
(4) M-theory on $G_2$ manifolds~\cite{Acharya:2001gy, Friedmann:2002ty}. 
Those models can be dual
to the heterotic models on Calabi-Yau threefolds or
to some Type II orientifold models.
(5) F-theory GUTs~\cite{Beasley:2008dc, Beasley:2008kw, Donagi:2008ca, Donagi:2008kj, Jiang:2009zza, Jiang:2009za}. The $SU(5)$ gauge symmetry can be broken
down to the SM gauge symmetries by turning on the $U(1)_Y$
fluxes, and the $SO(10)$ gauge symmetry can be broken down
to the flipped $SU(5)\times U(1)_X$ gauge symmetries and
the $SU(3)_C\times SU(2)_L \times SU(2)_R \times U(1)_{B-L}$
gauge symmetries by turning on the $U(1)_X$ and $U(1)_{B-L}$
fluxes respectively.

To stabilize the moduli fields, the string theories with flux compactifications
have also been studied~\cite{Bousso:2000xa, Giddings:2001yu, Kachru:2003aw, Susskind:2003kw, Denef:2004ze, Denef:2004cf}, in which there intriguingly exist huge meta-stable flux vacua. For example, in the Type IIB theory with
RR and NSNS flux compactifications, the number of the meta-stable 
flux vacua can be of order $10^{500}$~\cite{Denef:2004dm,Denef:2007pq}.
With a weak anthropic principle, this may provide a solution to
the cosmological constant problem and could explain the gauge hierarchy
problem as well.

For our work here in this paper, we study only the flipped $SU(5)\times U(1)_X$ models, and we now shall provide a brief review
of the minimal flipped $SU(5)\times U(1)_X$ model~\cite{Barr:1981qv, Derendinger:1983aj, Antoniadis:1987dx}. 
The gauge group of the flipped $SU(5)$ model is
$SU(5)\times U(1)_{X}$, which can be embedded into $SO(10)$.
We define the generator $U(1)_{Y'}$ in $SU(5)$ as 
\begin{eqnarray} 
T_{\rm U(1)_{Y'}}={\rm diag} \left(-\frac{1}{3}, -\frac{1}{3}, -\frac{1}{3},
 \frac{1}{2},  \frac{1}{2} \right).
\label{u1yp}
\end{eqnarray}
The hypercharge is given by
\begin{eqnarray}
Q_{Y} = \frac{1}{5} \left( Q_{X}-Q_{Y'} \right).
\label{ycharge}
\end{eqnarray}
In addition, 
there are three families of SM fermions 
whose quantum numbers under the $SU(5)\times U(1)_{X}$ gauge group are
\begin{eqnarray}
F_i={\mathbf{(10, 1)}},~ {\bar f}_i={\mathbf{(\bar 5, -3)}},~
{\bar l}_i={\mathbf{(1, 5)}},
\label{smfermions}
\end{eqnarray}
where $i=1, 2, 3$. 

To break the GUT and electroweak gauge symmetries, we 
introduce two pairs of Higgs fields
\begin{eqnarray}
&H={\mathbf{(10, 1)}},~{\overline{H}}={\mathbf{({\overline{10}}, -1)}},& \\ \nonumber
&~h={\mathbf{(5, -2)}},~{\overline h}={\mathbf{({\bar {5}}, 2)}}.&
\label{Higgse1}
\end{eqnarray}
Interestingly, we can naturally solve the doublet-triplet splitting
 problem via the missing partner mechanism~\cite{Antoniadis:1987dx}, and then
the dimension five
proton decay from the colored Higgsino exchange can be
highly suppressed~\cite{Antoniadis:1987dx}.
The flipped $SU(5)\times U(1)_X$ models have been
constructed systematically in the free fermionic string 
constructions at Kac-Moody level one previously~\cite{Antoniadis:1987dx,Antoniadis:1987tv, Antoniadis:1988tt, Antoniadis:1989zy,Lopez:1992kg},
and in the  F-theory model building recently~\cite{Beasley:2008dc, Beasley:2008kw, Donagi:2008ca, Donagi:2008kj, Jiang:2009zza, Jiang:2009za}, and we represent the flipped $SU(5)\times U(1)_X$ models as one pillar of the foundation for ${\cal F}$-$SU(5)$ in Fig.~\ref{fig:MVII_MultiversePyramid}.
In the flipped $SU(5)\times U(1)_X$ models, there are two unification
scales: the $SU(3)_C\times SU(2)_L$ unification scale $M_{32}$ and
the $SU(5)\times U(1)_X$ unification scale $M_{\cal F}$.
To separate the $M_{32}$ and $M_{\cal F}$ scales
and obtain true string-scale gauge coupling unification in 
free fermionic string models~\cite{Jiang:2006hf, Lopez:1992kg} or
the decoupling scenario in F-theory models~\cite{Jiang:2009zza, Jiang:2009za},
we introduce vector-like particles which form complete
flipped $SU(5)\times U(1)_X$ multiplets, and we insert the vector particles and F-Theory as a second pillar in Fig.~\ref{fig:MVII_MultiversePyramid}, and also integrate their presence into Fig.~\ref{fig:MVII_FLandscape_Stars}.
In order to avoid the Landau pole
problem for the strong coupling constant, we can only introduce the
following two sets of vector-like particles around the TeV 
scale~\cite{Jiang:2006hf}
\begin{eqnarray}
&& Z1:  XF ={\mathbf{(10, 1)}}~,~
{\overline{XF}}={\mathbf{({\overline{10}}, -1)}}~;~\\
&& Z2: XF~,~{\overline{XF}}~,~Xl={\mathbf{(1, -5)}}~,~
{\overline{Xl}}={\mathbf{(1, 5)}}
~,~\,
\end{eqnarray}
where 
\begin{eqnarray}
{XF} ~\equiv~ (XQ,XD^c,XN^c)~,~~~{\overline{Xl}}_{\mathbf{(1, 5)}}\equiv XE^c ~.~\,
\end{eqnarray}
In the prior, $XQ$, $XD^c$, $XE^c$, $XN^c$ have the same quantum numbers as the
quark doublet, the right-handed down-type quark, charged lepton, and
neutrino, respectively. 
Such kind of the models have been constructed 
systematically in the F-theory model building locally and dubbed 
${\cal F}-SU(5)$ within that context~\cite{Jiang:2009zza, Jiang:2009za}.
In this paper, we only consider the flipped
$SU(5)\times U(1)_X$ models with 
$Z2$ set of vector-like particles.
The discussions for the models with 
$Z1$ set and heavy threshold corrections~\cite{Jiang:2009zza, Jiang:2009za}
are similar.

\section{Super No-Scale Supergravity}

We now turn to the third and final pillar of the ${\cal F}$-$SU(5)$ foundation in Fig.~\ref{fig:MVII_MultiversePyramid}, that of No-Scale supergravity. In the traditional framework, 
supersymmetry is broken in 
the hidden sector, and then its breaking effects are
mediated to the observable sector
via gravity or gauge interactions. In GUTs with
gravity mediated supersymmetry breaking, also known as the
minimal supergravity (mSUGRA) model, 
the supersymmetry breaking soft terms can be parameterized
by four universal parameters: the gaugino mass $M_{1/2}$,
scalar mass $M_0$, trilinear soft term $A$, and
the ratio of Higgs VEVs $\tan \beta$ at low energy,
plus the sign of the Higgs bilinear mass term $\mu$.
The $\mu$ term and its bilinear 
soft term $B_{\mu}$ are determined
by the $Z$-boson mass $M_Z$ and $\tan \beta$ after
the electroweak (EW) symmetry breaking.

To solve the cosmological constant
problem, No-Scale supergravity was proposed~\cite{Cremmer:1983bf,Ellis:1983sf, Ellis:1983ei, Ellis:1984bm, Lahanas:1986uc}. 
No-scale supergravity is defined as the subset of supergravity models
which satisfy the following three constraints~\cite{Cremmer:1983bf,Ellis:1983sf, Ellis:1983ei, Ellis:1984bm, Lahanas:1986uc}:
(i) The vacuum energy vanishes automatically due to the suitable
 K\"ahler potential; (ii) At the minimum of the scalar
potential, there are flat directions which leave the 
gravitino mass $M_{3/2}$ undetermined; (iii) The super-trace
quantity ${\rm Str} {\cal M}^2$ is zero at the minimum. Without this,
the large one-loop corrections would force $M_{3/2}$ to be either
zero or of Planck scale. A simple K\"ahler potential which
satisfies the first two conditions is
\begin{eqnarray} 
K &=& -3 {\rm ln}( T+\overline{T}-\sum_i \overline{\Phi}_i
\Phi_i)~,~
\label{NS-Kahler}
\end{eqnarray}
where $T$ is a modulus field and $\Phi_i$ are matter fields.
The third condition is model dependent and can always be satisfied in
principle~\cite{Ferrara:1994kg}. We emphasize that No-Scale
supergravity can be realized in the compactification
of the weakly coupled heterotic string 
theory~\cite{Witten:1985xb} and the
compactification of M-theory on S1/Z2 at the
 leading order~\cite{Li:1997sk}.

The scalar fields in the above
K\"ahler potential parameterize the coset space
$SU(N_C+1, 1)/(SU(N_C+1)\times U(1))$, where $N_C$ is the number
of matter fields. Analogous structures appear in the 
$N\ge 5$ extended supergravity theories~\cite{Cremmer:1979up}, for example,
$N_C=4$ for $N=5$, which can be realized in the compactifications
of string theory~\cite{Witten:1985xb, Li:1997sk}. 
The non-compact structure of the symmetry
implies that the potential is not only constant but actually
identical to zero. In fact, one can easily check that
the scalar potential is automatically positive semi-definite,
and has a flat direction along the $T$ field. It is interesting that
for the simple K\"ahler potential in Equation~(\ref{NS-Kahler}),
we obtain the simplest No-Scale boundary condition
$M_0=A=B_{\mu}=0$, while $M_{1/2}$ may be
non-zero at the unification scale,
allowing for low energy SUSY breaking.

The single relevant modulus field in the simplest 
string No-Scale supergravity is the K\"ahler
modulus $T$, a characteristic of the Calabi-Yau manifold,
the dilaton coupling being irrelevant.
The F-term of $T$ generates the gravitino mass $M_{3/2}$, 
which is proportionally equivalent to $M_{1/2}$.
Exploiting the simplest No-Scale boundary condition at $M_{\cal F}$ and 
running from high energy to low energy under the RGEs,
there can be a secondary minimization, or MM, of the minimum of the
Higgs potential $V_{\rm min}$ for the EWSB vacuum.
Since $V_{\rm min}$ depends on $M_{1/2}$, the gaugino mass $M_{1/2}$ is consequently 
dynamically determined by the equation $dV_{\rm min}/dM_{1/2}=0$,
aptly referred to as the Super No-Scale mechanism~\cite{Li:2010uu,Li:2011dw,Li:2011xu}. In this paper, we shall define the universe as the MM of the effective Higgs potential for a given set of input parameters.

\begin{figure*}[ht]
	\centering
	\includegraphics[width=1.00\textwidth]{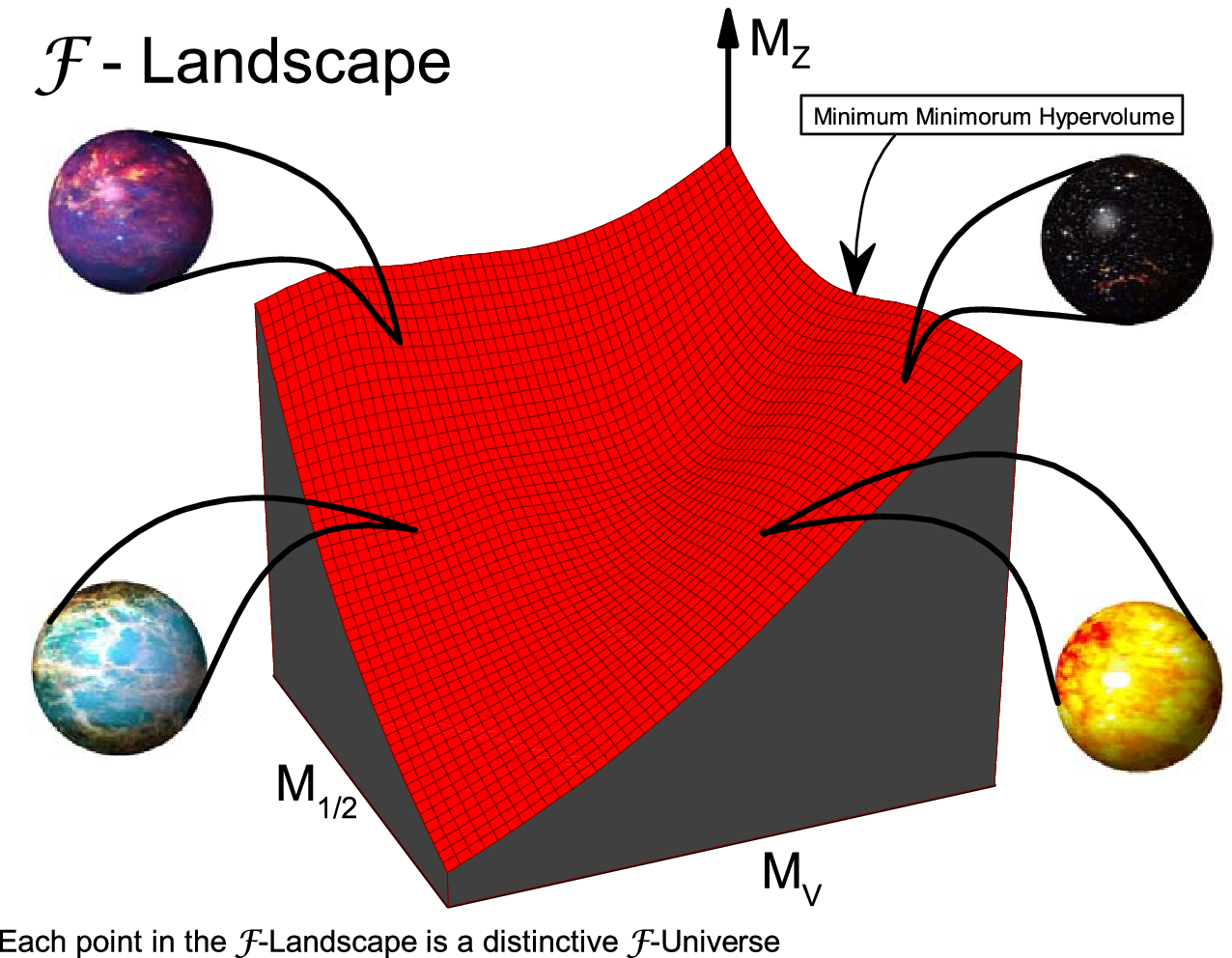}
	\caption{Heuristic depiction of the ${\cal F}$-Landscape in terms of the parameters $M_{1/2}, M_{V}$, and $M_{Z}$, with four indiscriminate universes shown at random points on the landscape. The hypervolume illustrated is a contiguous region of minimum minimorum, each minimum minimorum representing the minimum of the 1-loop scalar Higgs potential and an EWSB vacuum.}
	\label{fig:MVII_FlandscapeSpheres}
\end{figure*}

\section{Super No-Scale ${\cal F}$-$SU(5)$}

The model investigated here, dubbed No-Scale $\cal{F}$-$SU(5)$~\cite{Li:2010ws, Li:2010mi,Li:2010uu,Li:2011dw, Li:2011hr, Maxin:2011hy, Li:2011xu, Li:2011in,Li:2011gh,Li:2011rp,Li:2011fu,Li:2011xg}, unifies the ${\cal F}$-lipped $SU(5)$ Grand Unified Theory (GUT)~\cite{Barr:1981qv,Derendinger:1983aj,Antoniadis:1987dx} with
two pairs of hypothetical TeV-scale vector-like supersymmetric multiplets with origins in
${\cal F}$-theory~\cite{Jiang:2006hf,Jiang:2009zza,Jiang:2009za,Li:2010dp,Li:2010rz} and the dynamically established boundary conditions of No-Scale
supergravity~\cite{Cremmer:1983bf,Ellis:1983sf, Ellis:1983ei, Ellis:1984bm, Lahanas:1986uc}, as exhibited in Fig.~\ref{fig:MVII_MultiversePyramid}. A more complete review of this model is available in the appendix of Ref.~\cite{Maxin:2011hy}.

We have previously defined an exceptionally constrained Golden Point~\cite{Li:2010ws} and Golden Strip~\cite{Li:2010mi,Li:2011xg} using the dynamically established boundary conditions of No-Scale supergravity at the $\cal{F}$-$SU(5)$ unification scale that satisfied all the latest experimental constraints while also generating an imminently observable proton decay rate~\cite{Li:2009fq}. The most limiting constraint imposed upon the viable parameter space is the unification scale boundary on $B_{\mu}=0$. Moreover, the $M_{1/2}$ boundary gaugino mass and the ratio of the Higgs vacuum expectation values (VEVs) $\tan\beta$ were dynamically determined by applying the Super No-Scale condition for the dynamic stabilization of the stringy modulus related to $M_{1/2}$~\cite{Li:2010uu, Li:2011dw,Li:2011xu}.

The total set of supersymmetry breaking soft terms evolve from the single parameter $M_{1/2}$ in the simplest No-Scale supergravity, and as a result, the particle spectra are proportionally comparable up to an overall rescaling on $M_{1/2}$, leaving the majority of the ``internal'' 
physical properties invariant.  This rescaling capability on $M_{1/2}$ is atypical and has generally not been observed in dissimilar supersymmetry models, due to the presence of a large parameterization freedom, particularly with respect to a second independent boundary mass $M_0$ for scalar fields. The dependence on the vector-like mass parameter is rather weak, though this rescaling symmetry can be broken to a small degree by $M_V$.

If we temporarily fix the top quark mass $m_{\rm t}$ and vector-like mass $M_{\rm V}$, the dual EWSB minimization conditions first ascertain the Higgs bilinear mass term $\mu$ at $M_{\cal F}$, and secondly, since $B_\mu(M_{\cal F}) = 0$ has been previously fixed by the No-Scale boundary conditions,
establishes $\tan\beta$ as an implicit function of the universal gaugino boundary mass $M_{1/2}$. The minimum of the electroweak Higgs potential 
$(V_{EW})_{min}$ depends on $M_{1/2}$, and furthermore, the gaugino mass $M_{1/2}$ is related to the F-term of the modulus in string models,  hence the gaugino mass is determined by the equation $d(V_{EW})_{min}/dM_{1/2}=0$ due to the modulus stabilization~\cite{Ellis:1983sf,Lahanas:1986uc}.
At this locally smallest value of $V_{\rm min}(M_{1/2})$, which is the MM, the dynamic determination of $M_{1/2}$ is realized, and this is the Super No-Scale supergravity condition. 

\begin{figure*}[htp]
	\centering
	\includegraphics[width=0.95\textwidth]{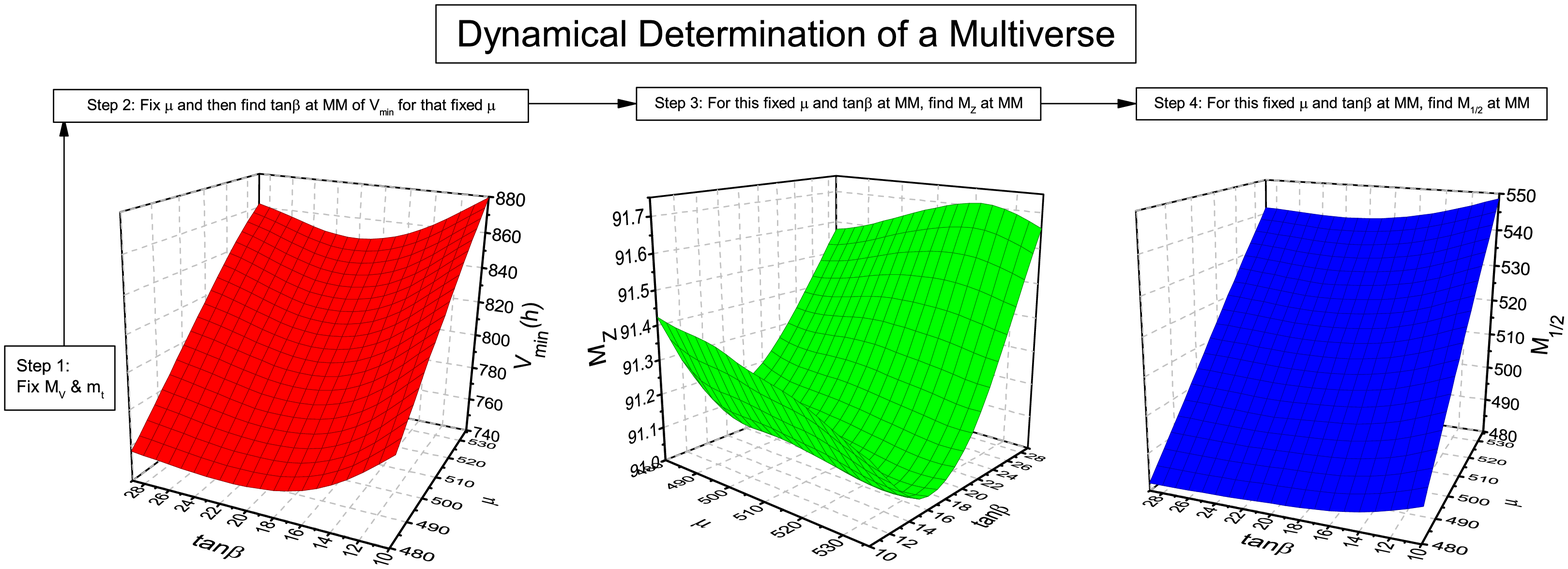}
	\includegraphics[width=0.95\textwidth]{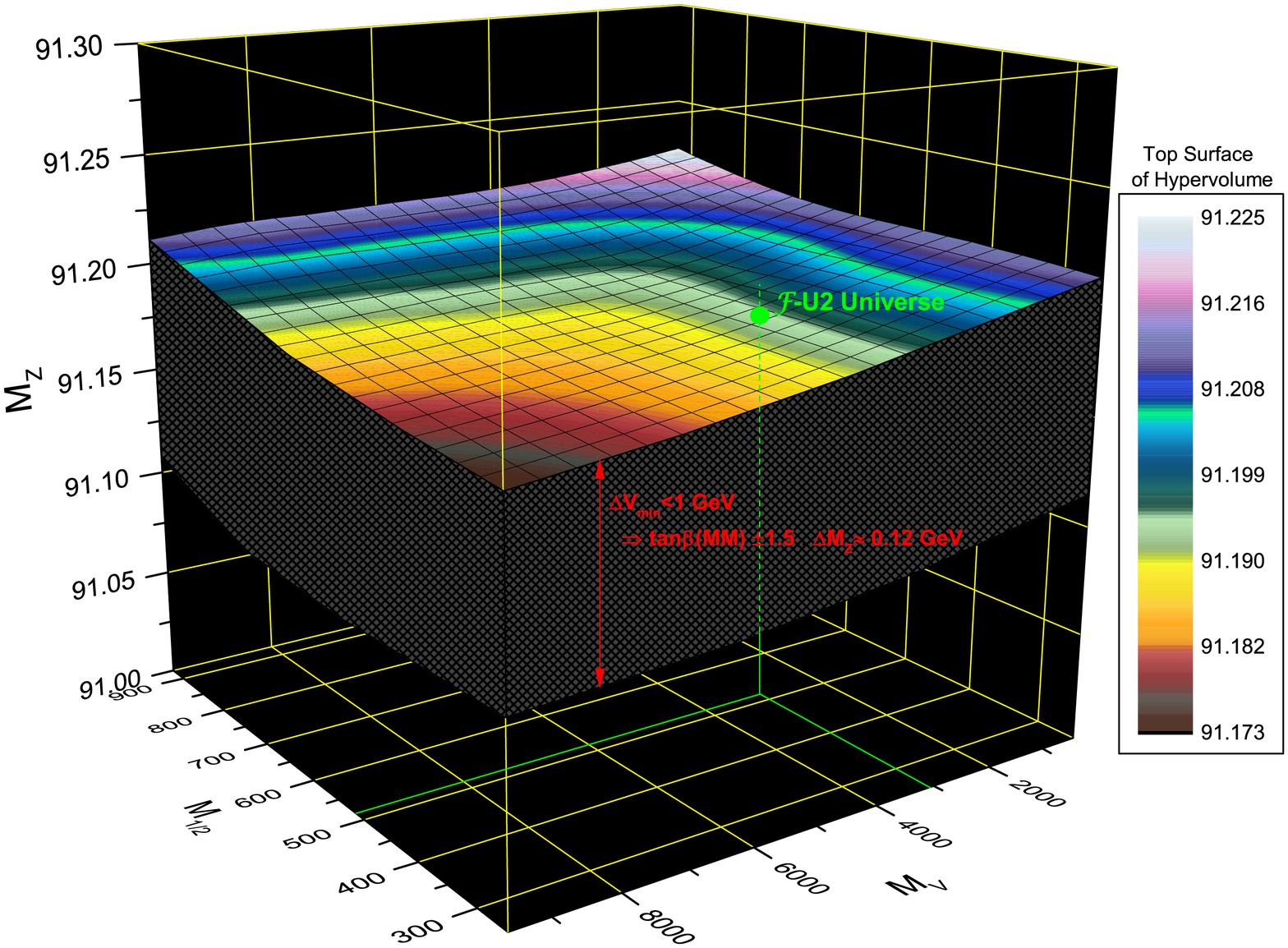}
	\caption{Logical flow depicting the process of dynamically determining a Multiverse. The upper three plot spaces show the $B_{\mu}$=0 hypersurfaces for a fixed set of ($M_V,m_t$). The lower plot space is generated by dynamically determining numerous points throughout the model space to estimate the hypervolume of minimum minimorum. Here we show the coordinates of universe ${\cal F}$-U2 in the lower plot from Table~\ref{tab:universes}, which represents a plausible candidate for our universe. The thickness of the volume is approximated by placing a constraint of $\Delta V_{min}(h) <$ 1 GeV at the minimum minimorum, similar in scale to the QCD corrections at the second loop. This results in a deviation of $\pm$1.5 on the tan$\beta$ at the minimum minimorum, translating into a 0.12 GeV uncertainty on the dynamically determined value of the electroweak scale $M_Z$.}
	\label{fig:MVII_Flowchart}
\end{figure*}

In this work we apply the expanded procedure implemented in~\cite{Li:2011dw, Li:2011xu}, where the
chief difference is that we now permit a fluctuation not just of $M_{1/2}$, which is related to the F-term of the
K\"ahler modulus $T$ in the weakly coupled heterotic $E_8 \times E_8$ string theory or in M-theory on
$S^1 / Z_2$, but of the GUT scale Higgs modulus as well, as represented in the mass scale $M_{32}$ at which
the $SU(3) \times SU(2)_{\rm L}$ couplings first unify. Using the low energy couplings as input, we could presume $M_{32}$ to be a ``given'' quantity.
In fact, beginning from the measured Standard Model gauge couplings and fermion Yukawa couplings at the LEP 
$91.187$~GeV electroweak scale, we can compute both $M_{32}$ and the final unification scale $M_{\cal F}$,
and consequently the unified gauge coupling and Standard Model fermion Yukawa couplings at $M_{\cal F}$, through
running of the RGEs.  Nonetheless, since the VEVs of the GUT Higgs fields $H$ and $\overline{H}$ are considered
here as free parameters, the GUT scale $M_{32}$ cannot be held constant either. As a result,
the low energy Standard Model gauge couplings, specifically the $SU(2)_L$ gauge coupling $g_2$,
will also run freely via this feedback from $M_{32}$. Upon realization of the existence of a second dynamic modulus, we do secure $\mu$ to a constant value, which being a basic numerical parameter, should be managed similarly to the top quark and vector-like mass parameters, and as such, a constant $\mu$ slice of the $V_{\rm min}$ hyper-surface is exacted out possessing a minimum value of the effective Higgs potential at the minimum of the parabolic curve. By removing the slice of fixed $\mu$ from the $V_{\rm min}$ hyper-surface, the secondary minimization condition on $\tan \beta$ is effectively rotated, albeit quite moderately, relative to the procedure of Ref.~(\cite{Li:2010uu}). The minimization we advocate here, referencing $M_{1/2}$, $M_{32}$ and $\tan \beta$, is again dependent upon $M_{V}$ and $m_{t}$, while the determination of $\tan \beta$ in~\cite{Li:2010uu}, by contrast, left $M_{V}$ and $m_{t}$ invariant. We emphasize the mutual consistency of the results, given the realization that an effective minimization requires modulation of all three parameters in order to assert a concurrent dynamical determination of all three parameters. It moreover confronts the difficulties of the SUSY breaking scale and gauge hierarchy~\cite{Li:2010uu}, insomuch as $M_{1/2}$ is determined dynamically.

Practically speaking, the variation of $M_{32}$ is accomplished in the reverse by a small programmatic variation of the Weinberg angle, with
$\sin^2 (\theta_{\rm W}) \simeq 0.236$, in excellent agreement with experiment, maintaining the strong and electromagnetic couplings at
their physically measured values. The magnitude of the Higgs VEV is in effect fixed, so the small shifting of the Weinberg Angle is achieved by a minor deviation in the $Z$-boson mass. Since the vital electroweak Higgs VEV is not a significant ingredient of the variation, we exercise caution when stating our claim for the dynamical determination of $M_Z \simeq 91.187$~GeV.  Although, in {\it conjunction} with the radiative electroweak
symmetry breaking~\cite{Ellis:1982wr, Ellis:1983bp} numerically implemented within the {\tt SuSpect 2.34} code base~\cite{Djouadi:2002ze},
the fixing of the Higgs VEV and the determination of the electroweak scale may also plausibly be considered 
legitimate dynamic output, {\it if} one posits the $M_{F}$ scale input to be available {\it a priori}. 

There is a pair of Higgs doublets
$H_u$ and $H_d$ which give mass to the up-type quarks and
down-type quarks/charged leptons, respectively, in a supersymmetric Standard Model. The one-loop effective
Higgs potential in the 't Hooft-Landau gauge and in the $\overline{\rm DR}$
scheme is given by
\begin{eqnarray}
V_{\rm eff} &=& V_0(H_u^0,~H_d^0) + V_1(H_u^0,~H_d^0)~,~\,
\end{eqnarray}
where 
\begin{eqnarray}
V_0&=& (\mu^2 + m_{H_u}^2) (H_u^0)^2 + (\mu^2 + m_{H_d}^2) (H_d^0)^2
\nonumber \\ &&
-2 B_{\mu} \mu H_u^0 H_d^0 + {\frac{g_2^2 + g_Y^2}{8}} \left[(H_u^0)^2-(H_d^0)^2\right]^2
 ~,~\,
\end{eqnarray}
\begin{eqnarray}
V_1 &=&  \sum_i {\frac{n_i}{64\pi^2}} m_i^4(\phi)
\left( {\rm ln}{{\frac{m_i^2(\phi)}{Q^2}} -{\frac{3}{2}}} \right) ~,~\,
\end{eqnarray}
where $m_{H_u}^2$ and $m_{H_d}^2$ are the supersymmetry breaking soft masses,
$g_2$ and $g_Y$ are respectively the gauge couplings of $SU(2)_L$ and
$U(1)_Y$, $n_i$ and $m_i^2(\phi)$ are respectively the degree of freedom and mass
for $\phi_i$, and $Q$ is the renormalization scale. We have revised the {\tt SuSpect 2.34} 
code base~\cite{Djouadi:2002ze} to incorporate our specialized
No-Scale ${\cal F}$-$SU(5)$ with vector-like mass algorithm, and appropriately implement
two-loop RGE running for the SM gauge couplings, and
one-loop RGE running for the SM fermion Yukawa couplings,
$\mu$ term, and SUSY breaking soft terms.

\section{The ${\cal F}$-Landscape}

\begin{figure*}[htp]
	\centering
	\includegraphics[width=0.90\textwidth]{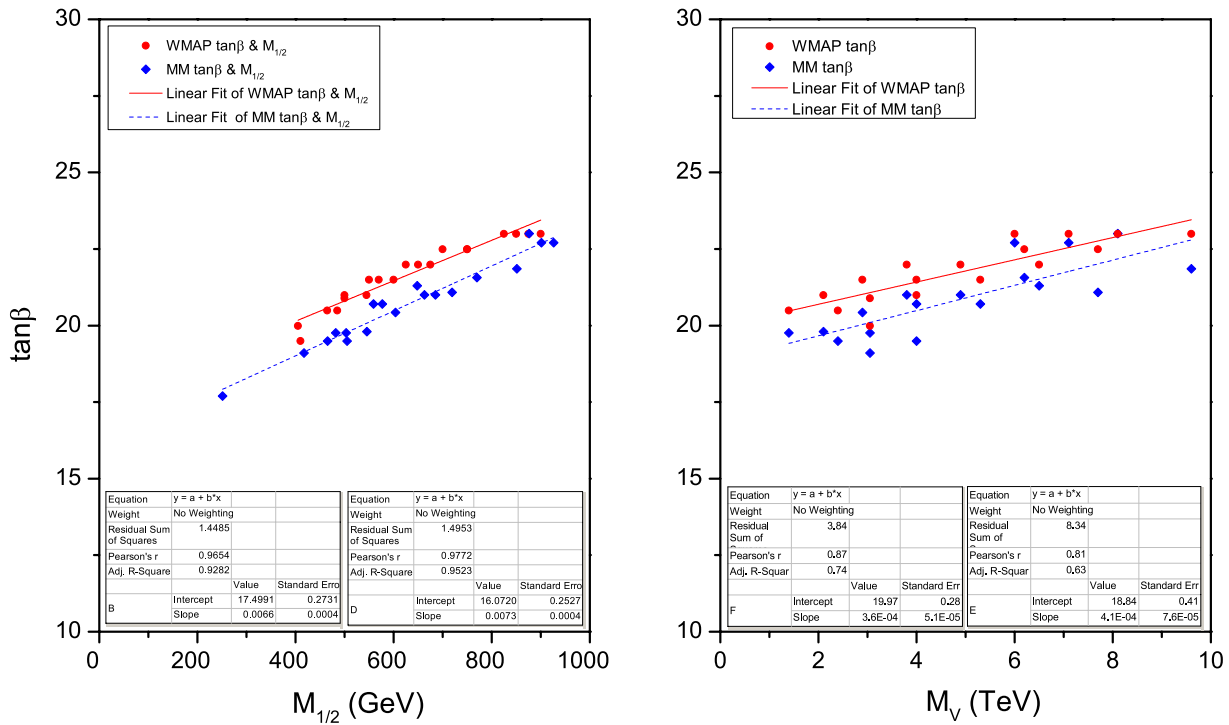}
	\caption{Scatter diagram comparing the tan$\beta$ necessary for generating the WMAP-7 relic density (circles) and the tan$\beta$ at the minimum minimorum (diamonds), for similar values of $M_{1/2}$ and $M_{V}$, in GeV. The linear fitted lines are nearly parallel and within the $\pm$1.5 constraint on tan$\beta$ at the minimum minimorum, indicating that the experimentally driven bottom-up strategy is self-consistent with the theoretically motivated top-down strategy, hinting at deep fundamental correlations within the model.}
	\label{fig:MVII_WMAPMMtanb}
\end{figure*}

The methodology of Refs.~\cite{Li:2011dw,Li:2011xu} for computing the MM summarized in the previous section had only been applied to a single point within the viable
${\cal F}$-$SU(5)$ parameter space in our previous work. We now seek to originate a full \textit{landscape} of the local ${\cal F}$-$SU(5)$ model space by calculating
the MM for a discrete set of points representative of the neighboring model space that presently subsists in the vicinity of the experimental uncertainties of our own universe.
Subsequently, we extrapolate the sampled findings to estimate a hypervolume of solutions for a more comprehensive panorama of the model space.  We shall then interpret this
landscape in the context of the \textit{Multiverse Blueprints}~\cite{Li:2011dw}, designating this subdivision as our local Multiverse community.  In a broader sense, the
Multiverse landscape is, of course, not limited to that zone which lies within our experimental uncertainty, though our purpose here is only initially to seek the prospective
structure of an ${\cal F}$-$SU(5)$ local Multiverse, within an acceptable introductory level of precision.  Each point within this No-Scale ${\cal F}$-$SU(5)$ Multiverse
landscape of solutions, which we shall heretofore refer to as the ${\cal F}$-Landscape, can be interpreted as a distinct universe within our regional dominion of universes,
as heuristically illustrated in Fig.~\ref{fig:MVII_FlandscapeSpheres}, where we portray four diverse universe samples active within the ${\cal F}$-Landscape.  The final
pane of Fig.~\ref{fig:MVII_FLandscape_Stars} demonstrates the resulting solution space, with application of rigorous numerics. As elaborated in~\cite{Li:2011dw}, testing of the
No-Scale ${\cal F}$-$SU(5)$ framework at the LHC is in some sense likewise a broader test of the framework of the String Landscape and the Multiverse of plausible string,
M- and F-theory vacua.  One can boldly speculate that substantiation of a No-Scale ${\cal F}$-$SU(5)$ configuration for our universe at the LHC offers indirect support for
a local dominion of ${\cal F}$-$SU(5)$ universes.

A modest sampling of satisfactory ${\cal F}$-$SU(5)$ points are extracted from the experimentally viable parameter space that satisfies the ``bare-minimal'' constraints
of~\cite{Li:2011xu}, in order to compute the MM in conformity with our conventional methodology summarized in the previous section. The outermost borders of the experimentally
viable parameter space presented in~\cite{Li:2011xu} are circumscribed from the bare-minimal constraints, though these constraints in principle are applicable only to our
universe and not the Multiverse in general. Nevertheless, the model space persisting within this constrained perimeter presents a generous supply of archetype universes to
explore and accordingly construct a hypervolume of solutions. To recapitulate, the bare-minimal constraints for our universe are defined by compatibility with the world
average top quark mass $m_{\rm t}$ = $173.3\pm 1.1$ GeV~\cite{:1900yx}, the prediction of a suitable candidate source of cold dark matter (CDM) relic density matching the upper and lower thresholds $0.1088 \leq \Omega_{CDM} \leq 0.1158$ set by the WMAP-7 measurements~\cite{Komatsu:2010fb}, a rigid prohibition against a charged lightest supersymmetric particle (LSP), compatibility with the precision LEP constraints on the lightest CP-even Higgs boson ($m_{h} \geq 114$ GeV~\cite{Barate:2003sz,Yao:2006px}) and other light SUSY chargino, stau, and neutralino mass content, and a self-consistency specification on the dynamically evolved value of $B_\mu$ measured at the boundary scale $M_{\cal{F}}$. An uncertainty of $\pm 1$~GeV on $B_\mu = 0$ is allowed, consistent with the induced variation from fluctuation of the strong coupling within its error bounds and the expected scale of radiative electroweak (EW) corrections. The lone constraint above that is necessarily mandatory for the Multiverse is that of the condition on the B-parameter at the $M_{\cal{F}}$ scale, since there is certainly no prerequisite for any of these other constrained parameters to inhabit within or even adjacent to the experimentally established uncertainties for our universe, although for our study here we prefer to remain nearby the local experimental ambiguities. The cumulative effect of these bare-minimal constraints distinctively shapes the experimentally viable parameter space germane to our universe into the uniquely formed profile situated in the ($M_{1/2},M_{\rm V}$) plane exhibited in Ref.~\cite{Li:2011xu}, from a tapered light mass region with a lower bound of $\tan \beta$ = 19.4 into a more expansive heavier region that ceases sharply with the charged stau LSP exclusion around tan$\beta \simeq$ 23. Correspondingly, we shall not journey too far afield from this narrow region of tan$\beta$ or the world average top quark periphery.

\begin{figure*}[ht]
	\centering
	\includegraphics[width=0.60\textwidth]{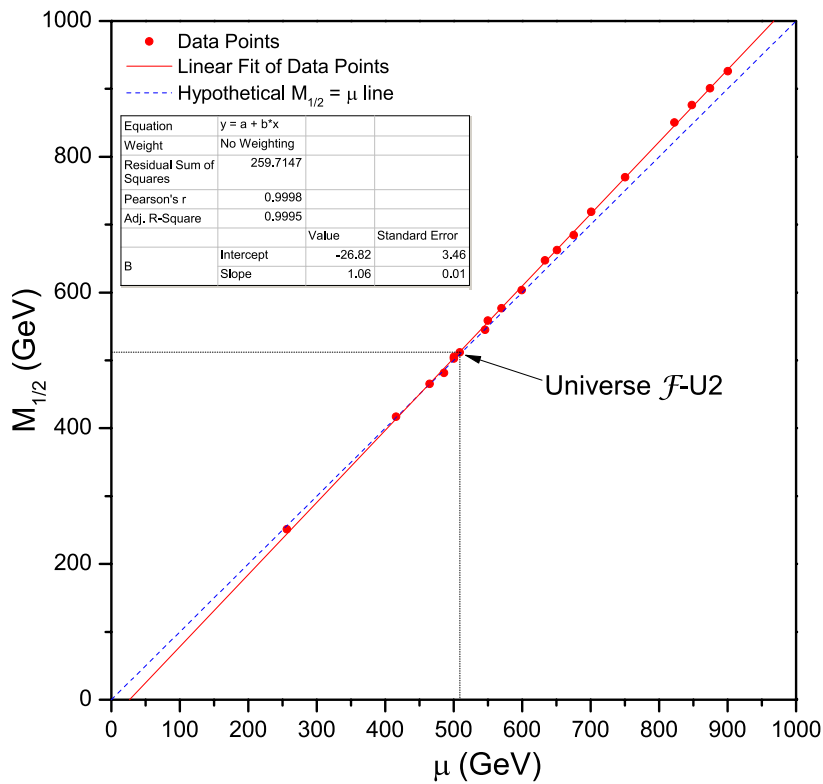}
	\caption{Data points depicting the linear relationship between $M_{1/2}$ and $\mu$, in GeV. Universe ${\cal F}$-U2, a phenomenologically favored candidate for our universe and also the most adept at explaining the CMS and ATLAS observations at the LHC, is highlighted for its noteworthy position right at the intersection of the data points linear fit and the hypothetical $M_{1/2}$ = $\mu$ line.}
	\label{fig:MVII_M12-mu}
\end{figure*}

The production of the hypervolume of solutions is initiated by mining the bare-minimally constrained wedge region in Ref.~\cite{Li:2011xu} for prospective universes from which to compute $V_{min}(h)$, carrying precision equivalent to the LEP constraints on the electroweak scale $M_Z$. In~\cite{Li:2011dw,Li:2011xu}, we executed the minimization procedure for a single specific fixed numerical value of $\mu$ only, so in essence, here we are broadening the blueprint of~\cite{Li:2011dw,Li:2011xu} to encompass an extensive range of $\mu$, utilizing our prescribed freedom of the numerical parameter. The secondary minimization procedure is thus enlarged by an order of cardinality, such that we may position the numerical value of $\mu$ to any figure we require, essentially dynamically determining in principle all $M_{1/2}$, tan$\beta$, and $M_{Z}$ for any preset permutation of $M_{V}$ and $m_{t}$.  This prescription can be replicated for an indefinite quantity of regional points within the model space in order to extrapolate the outcome to an estimated hypervolume comprising our local dominion of universes.  A logically sequenced rendering of the prescription for dynamically determining a Multiverse is illustrated in Fig.~\ref{fig:MVII_Flowchart}, with the top half of the Fig.~\ref{fig:MVII_Flowchart} space elucidating the minimization procedure for a unique predetermined duo of $M_{V}$ and $m_{t}$, while the bottom half of the plot space reveals a depiction of the conjectural hypervolume of universes.

Explicitly, the flow demonstrated in Fig.~\ref{fig:MVII_Flowchart}, after selection of a fixed combination of $M_{V}$ and $m_{t}$, proceeds first to pinpoint tan$\beta$ at the minimum of the 1-loop Higgs potential $V_{min}(h)$ for a precise numerical value of $\mu$, as depicted in the upper left element, which is now deemed the MM. The curved grid surfaces illustrated in the top half of the Fig.~\ref{fig:MVII_Flowchart} space characterize the hypersurface of $B_\mu = 0$ solutions. The effect of the $\pm 1$~GeV induced electroweak scale variations on the $B_\mu = 0$ condition translates into a small thickness of the $B_\mu = 0$ hypersurfaces in the top half of the Fig.~\ref{fig:MVII_Flowchart} space, though we suppress this in the diagrams here for simplicity. At first glance, tan$\beta$ at the MM appears to be constant in Fig.~\ref{fig:MVII_Flowchart}, though in fact it is not, as tan$\beta$ at the MM experiences a slight gradual continuous variation as the numerical value of $\mu$ is continuously adjusted. Once tan$\beta$ at the MM for our selection of $\mu$ is discovered, we can then resolve the corresponding $M_{Z}$ and $M_{1/2}$ at this MM by analyzing the center and right plots in the top half of the Fig.~\ref{fig:MVII_Flowchart} space. We have in no way up to this point deviated from the methodology of Refs.~\cite{Li:2011dw,Li:2011xu}. We have only demonstrated that guidelines established in Refs.~\cite{Li:2011dw,Li:2011xu} can be broadened to incorporate the selection of any $\mu$, such that the freedom on the bilinear $\mu$ parameter can in some sense be envisioned as a dial that can ``tune'' $M_{1/2}$, tan$\beta$, and $M_{Z}$ to that of any distinctive universe, for any and all prescribed sets of $M_{V}$ and $m_{t}$, traversing the $B_\mu = 0$ hypersurfaces.

\begin{table*}[htp]
	\centering
	\caption{Ten randomly selected benchmark universes, with the exception of universe ${\cal F}-U2$, with their model parameters and supersymmetry spectrum, in GeV. Universe ${\cal F}-U2$ was selected for its favorable phenomenological characteristics matching those of our universe, this being only one example of a universe that can dynamically determine our electroweak scale $M_Z$, though with an $M_{1/2}$ capable of generating LHC events in line with the actual data observations of the CMS and ATLAS Collaborations.}
		\begin{tabular}{|c||c|c|c|c|c|c|c||c|c|c|c|c|c|c|c|c|c|c|c|c|c|c|c|c|c|c|c|c|c|c|c|} \hline
${\cal F}-{\rm Universe}$&$M_{1/2}$&$M_{V}$&$m_{t}$&$tan\beta$&$M_{Z}$&$\mu$&$B_{\mu}$&$m_{\widetilde{\chi}_1^0}$&$m_{\widetilde{\chi}_2^0} / m_{\widetilde{\chi}_1^{\pm}}$&$m_{\widetilde{\tau}_1}$&$m_{\widetilde{e}_R}$&$m_{\widetilde{t}_1}$&$m_{\widetilde{b}_R}$&$m_{\widetilde{u}_R}$&$m_{\widetilde{d}_R}$&$m_{\widetilde{g}}$&$m_{h}$&$m_{A}$ \\	\hline
$	{\cal F}-{\rm U}1	$&$	482	$&$	1400	$&$	174.4	$&$	19.8	$&$	91.152	$&$	486	$&$	-0.22	$&$	91	$&$	198	$&$	107	$&$	183	$&$	514	$&$	878	$&$	990	$&$	1029	$&$	655	$&$	121.1	$&$	924	$ \\ \hline
$	{\cal F}-{\rm U}2	$&$	512	$&$	3050	$&$	173.1	$&$	21.0	$&$	91.187	$&$	509	$&$	0.39	$&$	101	$&$	218	$&$	110	$&$	193	$&$	554	$&$	906	$&$	1016	$&$	1054	$&$	705	$&$	120.0	$&$	923	$ \\ \hline
$	{\cal F}-{\rm U}3	$&$	512	$&$	4000	$&$	172.6	$&$	19.5	$&$	91.118	$&$	506	$&$	0.14	$&$	102	$&$	219	$&$	127	$&$	193	$&$	554	$&$	901	$&$	1003	$&$	1040	$&$	707	$&$	119.4	$&$	918	$ \\ \hline
$	{\cal F}-{\rm U}4	$&$	546	$&$	2100	$&$	174.3	$&$	19.8	$&$	91.145	$&$	545	$&$	-0.20	$&$	106	$&$	229	$&$	128	$&$	206	$&$	591	$&$	975	$&$	1092	$&$	1134	$&$	741	$&$	121.3	$&$	1011	$ \\ \hline
$	{\cal F}-{\rm U}5	$&$	577	$&$	4000	$&$	173.2	$&$	20.7	$&$	91.141	$&$	570	$&$	-0.75	$&$	116	$&$	249	$&$	134	$&$	217	$&$	631	$&$	1007	$&$	1123	$&$	1164	$&$	791	$&$	120.5	$&$	1017	$ \\ \hline
$	{\cal F}-{\rm U}6	$&$	604	$&$	2900	$&$	174.0	$&$	20.4	$&$	91.127	$&$	599	$&$	0.68	$&$	120	$&$	257	$&$	143	$&$	226	$&$	663	$&$	1062	$&$	1186	$&$	1229	$&$	820	$&$	121.5	$&$	1083	$ \\ \hline
$	{\cal F}-{\rm U}7	$&$	648	$&$	6500	$&$	172.8	$&$	21.3	$&$	91.143	$&$	633	$&$	0.86	$&$	134	$&$	286	$&$	153	$&$	242	$&$	716	$&$	1109	$&$	1231	$&$	1274	$&$	889	$&$	120.5	$&$	1097	$ \\ \hline
$	{\cal F}-{\rm U}8	$&$	719	$&$	7700	$&$	173.2	$&$	21.1	$&$	91.126	$&$	701	$&$	-0.88	$&$	151	$&$	320	$&$	177	$&$	268	$&$	797	$&$	1221	$&$	1350	$&$	1396	$&$	983	$&$	121.2	$&$	1203	$ \\ \hline
$	{\cal F}-{\rm U}9	$&$	877	$&$	8100	$&$	173.6	$&$	23.0	$&$	91.165	$&$	845	$&$	0.31	$&$	187	$&$	394	$&$	199	$&$	325	$&$	980	$&$	1466	$&$	1621	$&$	1676	$&$	1183	$&$	122.3	$&$	1414	$ \\ \hline
$	{\cal F}-{\rm U}10	$&$	901	$&$	7100	$&$	174.0	$&$	22.7	$&$	91.143	$&$	874	$&$	-0.54	$&$	192	$&$	404	$&$	206	$&$	334	$&$	1006	$&$	1512	$&$	1673	$&$	1730	$&$	1211	$&$	123.0	$&$	1469	$ \\ \hline
		\end{tabular}
		\label{tab:universes}
\end{table*}

The multistep minimization procedure is copied for a sizable quantity of points in the model space, generating the solution space in the lower half plot of Fig.~\ref{fig:MVII_Flowchart} through an extrapolation of the discrete returns. Only those sub one GeV perturbations about the minimum of the 1-loop Higgs potential are preserved, which we judge to be comparable in scale to the QCD corrections to the Higgs potential at the second loop. This constraint confines the value of tan$\beta$ at the MM to live within an expected $\pm$1.5 deviation around the absolute minimum of $V_{min}(h)$. Consequently, we can project the ensuing variation in $M_{Z}$ to be about $\pm$0.12 GeV at the MM. Thusly, over and above the freedom in $\mu$ to select different universes by ``tuning'' $M_{1/2}$, tan$\beta$, and $M_{Z}$ along a continuous string of MM, we must further recognize the indeterminate nature of these parameters at the MM from the QCD fluctuations providing some discretion on confinement of the MM to this theoretic one-dimensional string. Yet, it is essential to bear in mind that altering any one of these parameters will demand a compensating adjustment in one or more of the remaining parameters in order to transit along the $B_\mu = 0$ direction, engendering an additional unique point in the hypervolume of solutions, i.e. a unique universe in the Multiverse. These small fluctuations about the MM induce the diagrammed thickness of the hypervolume advertised in Fig.~\ref{fig:MVII_Flowchart}, where each singular point in the illustrated hypervolume exemplifies an individual universe in the Multiverse.

The points employed in the compilation of the $B_\mu = 0$ hypersurface and hypervolume of Multiverse solutions in Fig.~\ref{fig:MVII_Flowchart} were extracted from the experimentally viable parameter space delineated in Ref.~\cite{Li:2011xu}, where the contours of tan$\beta$ defining those regions consistent with the WMAP-7 relic density measurements progressively scale with both $M_{1/2}$ and $M_{V}$. As noted earlier, the WMAP-7 experimentally allowed parameter space spans from $\tan \beta$ = 19.4 to around tan$\beta \simeq$ 23, enveloping those regions of the model space regarded as credible contenders for our universe from a bottom-up experimental perspective. From a Multiverse frame of reference, the WMAP-7 region is extraneous, as any universe within the ${\cal F}$-Landscape may possess an intrinsic ``WMAP'' dark matter density, so to speak. In the process of dynamically determining the $M_{1/2}$, tan$\beta$, and $M_{Z}$ at the MM, relevant to the top-down theoretical perspective, there is little reason to anticipate (at least not from the point of view of an island universe) that the bottom-up and top-down techniques should be self-consistent at more than just a single point.
Nevertheless, this remarkable correspondence is unquestionably what is discovered, prompting curiosity at whether the correlation stems from a deep physical motivation.  In particular, the parallel transport of parameterization freedom exhibited by the phenomenological and dynamical treatments appears to support the conjectural application of this framework to a continuum of locally adjacent universes, each individually seated at its own dynamic MM.

A generous selection of points is plotted in Fig.~\ref{fig:MVII_WMAPMMtanb}, highlighting the $M_{1/2}$, $M_{V}$, and tan$\beta$ that can produce the WMAP-7 relic density, in conjunction with the fixing of tan$\beta$ at the absolute MM, given by a numerical value of $\mu$ which in turn correlates with an $M_{1/2}$ intimately resembling the WMAP-7 $M_{1/2}$. Note that the WMAP $M_{V}$ is equivalent to the $M_{V}$ at the MM since the vector mass is not a dynamically determined parameter. As visibly depicted in Fig.~\ref{fig:MVII_WMAPMMtanb}, the slope of the two linearly fitted lines are practically parallel, only displaced by a small delta on tan$\beta$, comfortably within our imposed $\pm$1.5 variance on tan$\beta$ from QCD 2-loop corrections. To recap, when scrutinizing only the precise MM, the tan$\beta$ at this exact MM scales nearly perfectly with the WMAP-7 tan$\beta$, or plainly stated, tan$\beta$(WMAP) = tan$\beta$(MM). That this extraordinary self-consistency should connect the experimentally and theoretically inspired strategies is by no means guaranteed, nor do we deem it to be accident. It epitomizes the multitude of profound correlations which have been observed amid our exhaustive exploration of the No-Scale ${\cal F}$-$SU(5)$ models.

A further noteworthy aspect emerging from the constitution of the Multiverse hypervolume is a rather suggestive linkage between the Higgs bilinear mass term $\mu$ and the gaugino mass $M_{1/2}$. With $\tan\beta$ constrained to the local vicinity of the 20 value through the phenomenological scheme, we wish to emphasize that this is consistent with the original motivation of No-Scale GUTs, since the Super No-Scale condition itself becomes quite subtle if the vector-like particle mass is much larger
than the sparticle masses~\cite{Cremmer:1983bf,Ellis:1983sf,Ellis:1983ei,Ellis:1984bm,Lahanas:1986uc}.
Interestingly, we find for the ${\cal F}$-Landscape that $M_{1/2}$ is virtually equal to $\mu$ across the entire region of the model space investigated here, as sketched in Fig.~\ref{fig:MVII_M12-mu}. This may be an effect of the strong No-Scale boundary conditions and might moreover have deep implications to the solution of the $\mu$ problem in the supersymmetric standard model~\cite{LMNW-P}. The fact that $\mu$ and $M_{\rm V}$ might be generated from the same mechanism~\cite{LMNW-P} represents an additional naturalness argument for the suggestion that
$\mu$ and $M_{\rm V}$ should be of the same order. Likewise as intriguing, Fig.~\ref{fig:MVII_M12-mu} indicates $M_{1/2}$ and $\mu$ are definitively equal in an abbreviated segment amidst the most phenomenologically preferred region of $M_{1/2}\sim$ 500 GeV.

\begin{figure*}[htp]
	\centering
	\includegraphics[width=0.75\textwidth]{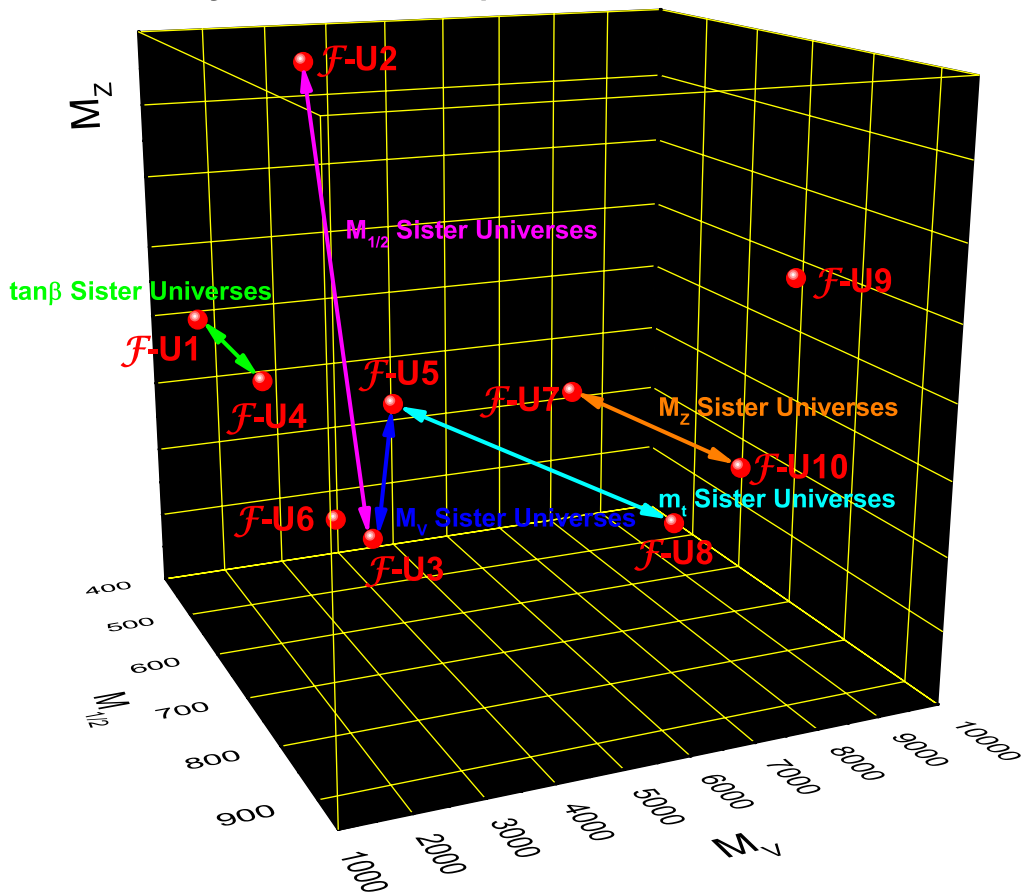}
	\caption{Three-dimensional diagram in ($M_{1/2},M_V,M_Z$) space of the ten benchmark universes of Table~\ref{tab:universes}, highlighting the Sister Universes of the these ten benchmarks. The Sister Universes are those that have an identical parameter to another universe. Each of the universes shown here are enclosed within the hypervolume of Fig.~\ref{fig:MVII_Flowchart}, though not shown here for clarity.}
	\label{fig:MVII_Sisters}
\end{figure*}

Ten benchmark universes are enumerated in Table~\ref{tab:universes}, introducing ${\cal F}$-U2 as a suitable contender for our universe, with the $(M_{1/2},M_V,M_Z)$ coordinates indigenous to the hypervolume annotated in Fig.~\ref{fig:MVII_Flowchart}. Universe ${\cal F}$-U2 possesses quite desirable phenomenological characteristics, deserving of its emphasis here. In particular, ${\cal F}$-U2 is emblematic of the class of universes endemic to the Golden Strip~\cite{Li:2011xg}, fulfilling all the most current experimental constraints to embody an exemplary target for supersymmetry discovery at the LHC. Moreover, the gaugino mass $M_{1/2}$ of universe ${\cal F}$-U2 adeptly explains compelling excesses in multijet events thus far observed by the CMS and ATLAS Collaborations at the LHC, producing categorical statistical fittings upon the genuine data event observations, discussed in~\cite{LMNW-P}.

Evidently from Fig.~\ref{fig:MVII_Flowchart}, distinct universes will nonetheless enjoy complementary facets as evidenced by their equivalent ${\cal F}$-$SU(5)$ model parameters. For instance, to borrow a traditional metaphor for associating companion entities, in Fig.~\ref{fig:MVII_Sisters} we discern that universes with invariant $M_{V}$ can be promulgated as $\textit{sister universes}$ in the vector particle mass. In the same vein, we can advocate sister universes for all the model parameters.  Presenting an explicit example, in the benchmark universes displayed in Fig.~\ref{fig:MVII_Sisters}, we display one conceivable sister universe of ${\cal F}$-U2 out of an innumerable distribution of sisters, that being the $M_{1/2}$ sister universe ${\cal F}$-U3. Insomuch as universe ${\cal F}$-U3 will brandish different $M_{V}$, $m_t$, tan$\beta$, and $M_Z$, it will feature a common gaugino mass with universe ${\cal F}$-2. The effect the sister attributes will imprint upon the tangible architecture of each universe is unknown, though one could lucidly speculate that sister parameters could translate into some analogous physical properties.


\section{Conclusions}

Modern Science has observed radical advancements in our conceptions of the cosmos as we peer deeper into the ``heavens'', both immeasurably far above and sub-microscopically small below. One of the most captivating conjectural notions in recent memory is the suggestion of a Multiverse, capable of resolving persistent dilemmas regarding the apparently finely-tuned physical properties of our universe. Nevertheless, historically the concept of a Multiverse has represented a generic solution with no model building aspects attached to it, since there existed no effective model capable of building the universes pervading a legitimate testable framework.  We have introduced here a analytical attempt at ``universe building'', constructing the Multiverse from a specific fundamental high-energy theory capable of describing our observable low-energy physics.

The framework we apply for our local dominion of universes is the model dubbed ${\cal F}$-$SU(5)$, which as we have shown here and in our numerous previous explorations, has proven an extraordinary consistency amongst parameters dynamically determined (the top-down approach) and parameters evaluated through application of the latest experimental constraints (the bottom-up approach).
Forged upon the foundations of the flipped $SU(5)$ GUT, extra TeV-Scale vector-like multiplets derived out of F-Theory, and the dynamics of No-Scale supergravity, ${\cal F}$-$SU(5)$ has not only persevered in the face of rapidly progressing constraints imposed by the LHC (while most alternative supersymmetric models such as mSUGRA and the CMSSM have been decimated), but ${\cal F}$-$SU(5)$ also moreover cleanly explains recent low-statistics excesses in multijet observations at the LHC.  Hence, the ${\cal F}$-$SU(5)$ is demonstrating itself to be a credible contender for the supersymmetric GUT for our universe. We suggest that experimental substantiation of a No-Scale ${\cal F}$-$SU(5)$ composition for our universe will thus enhance the case for string, M-, and F-Theory as a master theory of the Multiverse, with a ubiquitous No-Scale structure. As such, we have here constructed a locally phenomenologically adjacent sector of the Multiverse from our ${\cal F}$-$SU(5)$ M(ultiverse)-Theory.

By dynamically determining numerous points within the model space, we assembled a hypervolume of minimum minimorum, each point representing a distinctive EWSB vacuum capable of describing a unique universe. A viable candidate for our universe was presented that thus far nicely matches our observable phenomenology, and we offered up potential sister universes we may have inherited. Curiously, the fine threads of our journey into the Multiverse revealed two profound fundamental correlations, that of $M_{1/2}$=$\mu$ and tan$\beta$(WMAP)=tan$\beta$(MM). Any direct correspondence between the gaugino mass and the $\mu$ parameter could have significant ramifications toward a solution to the $\mu$ problem in the supersymmetric Standard Model, hence we continue to closely study this most interesting correlation. The surprising connection between disparate approaches to the derivations of tan$\beta$ at low-energy is of equal significance, insinuating an inseparable linkage between physically measured observables and the theoretical dynamics of ${\cal F}$-$SU(5)$.

We have come historically to fully expect new revelations that will continue to challenge that which our imaginations can perceive. Though the notion of a Multiverse
initially boggles the mind, as with many striking new developments in science and technology, such innovative conceptions may eventually gravitate from bizarre to canonical.
We anticipate that as the Multiverse framework continues to be rationally probed by sound principles of physics,
the idea that our universe is but one among an innumerable host will come to seem not so outlandish after all.


\begin{acknowledgments}
This research was supported in part 
by the DOE grant DE-FG03-95-Er-40917 (TL and DVN),
by the Natural Science Foundation of China 
under grant numbers 10821504 and 11075194 (TL),
by the Mitchell-Heep Chair in High Energy Physics (JAM),
and by the Sam Houston State University
2011 Enhancement Research Grant program (JWW).
We also thank Sam Houston State University
for providing high performance computing resources.
\end{acknowledgments}


\bibliography{bibliography}

\begin{thebibliography}{81}
\expandafter\ifx\csname natexlab\endcsname\relax\def\natexlab#1{#1}\fi
\expandafter\ifx\csname bibnamefont\endcsname\relax
  \def\bibnamefont#1{#1}\fi
\expandafter\ifx\csname bibfnamefont\endcsname\relax
  \def\bibfnamefont#1{#1}\fi
\expandafter\ifx\csname citenamefont\endcsname\relax
  \def\citenamefont#1{#1}\fi
\expandafter\ifx\csname url\endcsname\relax
  \def\url#1{\texttt{#1}}\fi
\expandafter\ifx\csname urlprefix\endcsname\relax\def\urlprefix{URL }\fi
\providecommand{\bibinfo}[2]{#2}
\providecommand{\eprint}[2][]{\url{#2}}

\bibitem[{\citenamefont{Denef et~al.}(2004)\citenamefont{Denef, Douglas, and
  Florea}}]{Denef:2004dm}
\bibinfo{author}{\bibfnamefont{F.}~\bibnamefont{Denef}},
  \bibinfo{author}{\bibfnamefont{M.~R.} \bibnamefont{Douglas}},
  \bibnamefont{and} \bibinfo{author}{\bibfnamefont{B.}~\bibnamefont{Florea}},
  {``}\bibinfo{title}{{Building a better racetrack}},{''}
  \bibinfo{journal}{JHEP} \textbf{\bibinfo{volume}{0406}}, \bibinfo{pages}{034}
  (\bibinfo{year}{2004}), \eprint{hep-th/0404257}.

\bibitem[{\citenamefont{Denef et~al.}(2007)\citenamefont{Denef, Douglas, and
  Kachru}}]{Denef:2007pq}
\bibinfo{author}{\bibfnamefont{F.}~\bibnamefont{Denef}},
  \bibinfo{author}{\bibfnamefont{M.~R.} \bibnamefont{Douglas}},
  \bibnamefont{and} \bibinfo{author}{\bibfnamefont{S.}~\bibnamefont{Kachru}},
  {``}\bibinfo{title}{{Physics of String Flux Compactifications}},{''}
  \bibinfo{journal}{Ann. Rev.Nucl.Part.Sci.} \textbf{\bibinfo{volume}{57}},
  \bibinfo{pages}{119} (\bibinfo{year}{2007}), \eprint{hep-th/0701050}.

\bibitem[{\citenamefont{Li et~al.}(2011{\natexlab{a}})\citenamefont{Li, Maxin,
  Nanopoulos, and Walker}}]{Li:2011dw}
\bibinfo{author}{\bibfnamefont{T.}~\bibnamefont{Li}},
  \bibinfo{author}{\bibfnamefont{J.~A.} \bibnamefont{Maxin}},
  \bibinfo{author}{\bibfnamefont{D.~V.} \bibnamefont{Nanopoulos}},
  \bibnamefont{and} \bibinfo{author}{\bibfnamefont{J.~W.}
  \bibnamefont{Walker}}, {``}\bibinfo{title}{{Blueprints of the No-Scale
  Multiverse at the LHC}},{''} \bibinfo{journal}{Phys. Rev.}
  \textbf{\bibinfo{volume}{D84}}, \bibinfo{pages}{056016}
  (\bibinfo{year}{2011}{\natexlab{a}}), \eprint{1101.2197}.

\bibitem[{\citenamefont{Li et~al.}(2011{\natexlab{b}})\citenamefont{Li, Maxin,
  Nanopoulos, and Walker}}]{Li:2010uu}
\bibinfo{author}{\bibfnamefont{T.}~\bibnamefont{Li}},
  \bibinfo{author}{\bibfnamefont{J.~A.} \bibnamefont{Maxin}},
  \bibinfo{author}{\bibfnamefont{D.~V.} \bibnamefont{Nanopoulos}},
  \bibnamefont{and} \bibinfo{author}{\bibfnamefont{J.~W.}
  \bibnamefont{Walker}}, {``}\bibinfo{title}{{Super No-Scale ${\cal
  F}$-$SU(5)$: Resolving the Gauge Hierarchy Problem by Dynamic Determination
  of $M_{1/2}$ and $\tan\beta$}},{''} \bibinfo{journal}{Phys. Lett. B}
  \textbf{\bibinfo{volume}{703}}, \bibinfo{pages}{469}
  (\bibinfo{year}{2011}{\natexlab{b}}), \eprint{1010.4550}.

\bibitem[{\citenamefont{Li et~al.}(2011{\natexlab{c}})\citenamefont{Li, Maxin,
  Nanopoulos, and Walker}}]{Li:2011xu}
\bibinfo{author}{\bibfnamefont{T.}~\bibnamefont{Li}},
  \bibinfo{author}{\bibfnamefont{J.~A.} \bibnamefont{Maxin}},
  \bibinfo{author}{\bibfnamefont{D.~V.} \bibnamefont{Nanopoulos}},
  \bibnamefont{and} \bibinfo{author}{\bibfnamefont{J.~W.}
  \bibnamefont{Walker}}, {``}\bibinfo{title}{{The Unification of Dynamical
  Determination and Bare Minimal Phenomenological Constraints in No-Scale
  \cal{F}- SU(5)}},{''} (\bibinfo{year}{2011}{\natexlab{c}}),
  \eprint{1105.3988}.

\bibitem[{\citenamefont{Witten}(1995)}]{Witten:1995ex}
\bibinfo{author}{\bibfnamefont{E.}~\bibnamefont{Witten}},
  {``}\bibinfo{title}{{String theory dynamics in various dimensions}},{''}
  \bibinfo{journal}{Nucl.Phys.} \textbf{\bibinfo{volume}{B443}},
  \bibinfo{pages}{85} (\bibinfo{year}{1995}), \eprint{hep-th/9503124}.

\bibitem[{\citenamefont{Vafa}(1996)}]{Vafa:1996xn}
\bibinfo{author}{\bibfnamefont{C.}~\bibnamefont{Vafa}},
  {``}\bibinfo{title}{{Evidence for F theory}},{''}
  \bibinfo{journal}{Nucl.Phys.} \textbf{\bibinfo{volume}{B469}},
  \bibinfo{pages}{403} (\bibinfo{year}{1996}), \eprint{hep-th/9602022}.

\bibitem[{\citenamefont{Polchinski}(1995)}]{Polchinski:1995mt}
\bibinfo{author}{\bibfnamefont{J.}~\bibnamefont{Polchinski}},
  {``}\bibinfo{title}{{Dirichlet Branes and Ramond-Ramond charges}},{''}
  \bibinfo{journal}{Phys.Rev.Lett.} \textbf{\bibinfo{volume}{75}},
  \bibinfo{pages}{4724} (\bibinfo{year}{1995}), \eprint{hep-th/9510017}.

\bibitem[{\citenamefont{Buchmuller et~al.}(2006)\citenamefont{Buchmuller,
  Hamaguchi, Lebedev, and Ratz}}]{Buchmuller:2005jr}
\bibinfo{author}{\bibfnamefont{W.}~\bibnamefont{Buchmuller}},
  \bibinfo{author}{\bibfnamefont{K.}~\bibnamefont{Hamaguchi}},
  \bibinfo{author}{\bibfnamefont{O.}~\bibnamefont{Lebedev}}, \bibnamefont{and}
  \bibinfo{author}{\bibfnamefont{M.}~\bibnamefont{Ratz}},
  {``}\bibinfo{title}{{Supersymmetric standard model from the heterotic
  string}},{''} \bibinfo{journal}{Phys.Rev.Lett.}
  \textbf{\bibinfo{volume}{96}}, \bibinfo{pages}{121602}
  (\bibinfo{year}{2006}), \eprint{hep-ph/0511035}.

\bibitem[{\citenamefont{Lebedev et~al.}(2007)\citenamefont{Lebedev, Nilles,
  Raby, Ramos-Sanchez, Ratz et~al.}}]{Lebedev:2006kn}
\bibinfo{author}{\bibfnamefont{O.}~\bibnamefont{Lebedev}},
  \bibinfo{author}{\bibfnamefont{H.~P.} \bibnamefont{Nilles}},
  \bibinfo{author}{\bibfnamefont{S.}~\bibnamefont{Raby}},
  \bibinfo{author}{\bibfnamefont{S.}~\bibnamefont{Ramos-Sanchez}},
  \bibinfo{author}{\bibfnamefont{M.}~\bibnamefont{Ratz}}, \bibnamefont{et~al.},
  {``}\bibinfo{title}{{A Mini-landscape of exact MSSM spectra in heterotic
  orbifolds}},{''} \bibinfo{journal}{Phys.Lett.}
  \textbf{\bibinfo{volume}{B645}}, \bibinfo{pages}{88} (\bibinfo{year}{2007}),
  \eprint{hep-th/0611095}.

\bibitem[{\citenamefont{Kim and Kyae}(2007)}]{Kim:2006hw}
\bibinfo{author}{\bibfnamefont{J.~E.} \bibnamefont{Kim}} \bibnamefont{and}
  \bibinfo{author}{\bibfnamefont{B.}~\bibnamefont{Kyae}},
  {``}\bibinfo{title}{{Flipped SU(5) from Z(12-I) orbifold with Wilson
  line}},{''} \bibinfo{journal}{Nucl.Phys.} \textbf{\bibinfo{volume}{B770}},
  \bibinfo{pages}{47} (\bibinfo{year}{2007}), \eprint{hep-th/0608086}.

\bibitem[{\citenamefont{Braun et~al.}(2005)\citenamefont{Braun, He, Ovrut, and
  Pantev}}]{Braun:2005ux}
\bibinfo{author}{\bibfnamefont{V.}~\bibnamefont{Braun}},
  \bibinfo{author}{\bibfnamefont{Y.-H.} \bibnamefont{He}},
  \bibinfo{author}{\bibfnamefont{B.~A.} \bibnamefont{Ovrut}}, \bibnamefont{and}
  \bibinfo{author}{\bibfnamefont{T.}~\bibnamefont{Pantev}},
  {``}\bibinfo{title}{{A Heterotic standard model}},{''}
  \bibinfo{journal}{Phys.Lett.} \textbf{\bibinfo{volume}{B618}},
  \bibinfo{pages}{252} (\bibinfo{year}{2005}), \eprint{hep-th/0501070}.

\bibitem[{\citenamefont{Bouchard and Donagi}(2006)}]{Bouchard:2005ag}
\bibinfo{author}{\bibfnamefont{V.}~\bibnamefont{Bouchard}} \bibnamefont{and}
  \bibinfo{author}{\bibfnamefont{R.}~\bibnamefont{Donagi}},
  {``}\bibinfo{title}{{An SU(5) heterotic standard model}},{''}
  \bibinfo{journal}{Phys.Lett.} \textbf{\bibinfo{volume}{B633}},
  \bibinfo{pages}{783} (\bibinfo{year}{2006}), \eprint{hep-th/0512149}.

\bibitem[{\citenamefont{Antoniadis
  et~al.}(1988{\natexlab{a}})\citenamefont{Antoniadis, Ellis, Hagelin, and
  Nanopoulos}}]{Antoniadis:1987tv}
\bibinfo{author}{\bibfnamefont{I.}~\bibnamefont{Antoniadis}},
  \bibinfo{author}{\bibfnamefont{J.~R.} \bibnamefont{Ellis}},
  \bibinfo{author}{\bibfnamefont{J.}~\bibnamefont{Hagelin}}, \bibnamefont{and}
  \bibinfo{author}{\bibfnamefont{D.~V.} \bibnamefont{Nanopoulos}},
  {``}\bibinfo{title}{{GUT Model Building with Fermionic Four-Dimensional
  Strings}},{''} \bibinfo{journal}{Phys.Lett.} \textbf{\bibinfo{volume}{B205}},
  \bibinfo{pages}{459} (\bibinfo{year}{1988}{\natexlab{a}}).

\bibitem[{\citenamefont{Antoniadis
  et~al.}(1988{\natexlab{b}})\citenamefont{Antoniadis, Ellis, Hagelin, and
  Nanopoulos}}]{Antoniadis:1988tt}
\bibinfo{author}{\bibfnamefont{I.}~\bibnamefont{Antoniadis}},
  \bibinfo{author}{\bibfnamefont{J.~R.} \bibnamefont{Ellis}},
  \bibinfo{author}{\bibfnamefont{J.~S.} \bibnamefont{Hagelin}},
  \bibnamefont{and} \bibinfo{author}{\bibfnamefont{D.~V.}
  \bibnamefont{Nanopoulos}}, {``}\bibinfo{title}{{An Improved SU(5) x U(1)
  Model from Four-Dimensional String}},{''} \bibinfo{journal}{Phys.Lett.}
  \textbf{\bibinfo{volume}{B208}}, \bibinfo{pages}{209}
  (\bibinfo{year}{1988}{\natexlab{b}}).

\bibitem[{\citenamefont{Antoniadis et~al.}(1989)\citenamefont{Antoniadis,
  Ellis, Hagelin, and Nanopoulos}}]{Antoniadis:1989zy}
\bibinfo{author}{\bibfnamefont{I.}~\bibnamefont{Antoniadis}},
  \bibinfo{author}{\bibfnamefont{J.~R.} \bibnamefont{Ellis}},
  \bibinfo{author}{\bibfnamefont{J.}~\bibnamefont{Hagelin}}, \bibnamefont{and}
  \bibinfo{author}{\bibfnamefont{D.~V.} \bibnamefont{Nanopoulos}},
  {``}\bibinfo{title}{{The Flipped SU(5) x U(1) String Model Revamped}},{''}
  \bibinfo{journal}{Phys.Lett.} \textbf{\bibinfo{volume}{B231}},
  \bibinfo{pages}{65} (\bibinfo{year}{1989}).

\bibitem[{\citenamefont{Faraggi et~al.}(1990)\citenamefont{Faraggi, Nanopoulos,
  and Yuan}}]{Faraggi:1989ka}
\bibinfo{author}{\bibfnamefont{A.~E.} \bibnamefont{Faraggi}},
  \bibinfo{author}{\bibfnamefont{D.~V.} \bibnamefont{Nanopoulos}},
  \bibnamefont{and} \bibinfo{author}{\bibfnamefont{K.-j.} \bibnamefont{Yuan}},
  {``}\bibinfo{title}{{A Standard Like Model in the 4D Free Fermionic String
  Formulation}},{''} \bibinfo{journal}{Nucl.Phys.}
  \textbf{\bibinfo{volume}{B335}}, \bibinfo{pages}{347} (\bibinfo{year}{1990}).

\bibitem[{\citenamefont{Antoniadis et~al.}(1990)\citenamefont{Antoniadis,
  Leontaris, and Rizos}}]{Antoniadis:1990hb}
\bibinfo{author}{\bibfnamefont{I.}~\bibnamefont{Antoniadis}},
  \bibinfo{author}{\bibfnamefont{G.}~\bibnamefont{Leontaris}},
  \bibnamefont{and} \bibinfo{author}{\bibfnamefont{J.}~\bibnamefont{Rizos}},
  {``}\bibinfo{title}{{A Three generation SU(4) x O(4) string model}},{''}
  \bibinfo{journal}{Phys.Lett.} \textbf{\bibinfo{volume}{B245}},
  \bibinfo{pages}{161} (\bibinfo{year}{1990}).

\bibitem[{\citenamefont{Lopez et~al.}(1993)\citenamefont{Lopez, Nanopoulos, and
  Yuan}}]{Lopez:1992kg}
\bibinfo{author}{\bibfnamefont{J.~L.} \bibnamefont{Lopez}},
  \bibinfo{author}{\bibfnamefont{D.~V.} \bibnamefont{Nanopoulos}},
  \bibnamefont{and} \bibinfo{author}{\bibfnamefont{K.-j.} \bibnamefont{Yuan}},
  {``}\bibinfo{title}{{The Search for a realistic flipped SU(5) string
  model}},{''} \bibinfo{journal}{Nucl.Phys.} \textbf{\bibinfo{volume}{B399}},
  \bibinfo{pages}{654} (\bibinfo{year}{1993}), \eprint{hep-th/9203025}.

\bibitem[{\citenamefont{Cleaver et~al.}(2002)\citenamefont{Cleaver, Faraggi,
  Nanopoulos, and Walker}}]{Cleaver:2001ab}
\bibinfo{author}{\bibfnamefont{G.}~\bibnamefont{Cleaver}},
  \bibinfo{author}{\bibfnamefont{A.}~\bibnamefont{Faraggi}},
  \bibinfo{author}{\bibfnamefont{D.~V.} \bibnamefont{Nanopoulos}},
  \bibnamefont{and} \bibinfo{author}{\bibfnamefont{J.}~\bibnamefont{Walker}},
  {``}\bibinfo{title}{{Phenomenology of nonAbelian flat directions in a minimal
  superstring standard model}},{''} \bibinfo{journal}{Nucl.Phys.}
  \textbf{\bibinfo{volume}{B620}}, \bibinfo{pages}{259} (\bibinfo{year}{2002}),
  \eprint{hep-ph/0104091}.

\bibitem[{\citenamefont{Berkooz et~al.}(1996)\citenamefont{Berkooz, Douglas,
  and Leigh}}]{Berkooz:1996km}
\bibinfo{author}{\bibfnamefont{M.}~\bibnamefont{Berkooz}},
  \bibinfo{author}{\bibfnamefont{M.~R.} \bibnamefont{Douglas}},
  \bibnamefont{and} \bibinfo{author}{\bibfnamefont{R.~G.} \bibnamefont{Leigh}},
  {``}\bibinfo{title}{{Branes intersecting at angles}},{''}
  \bibinfo{journal}{Nucl.Phys.} \textbf{\bibinfo{volume}{B480}},
  \bibinfo{pages}{265} (\bibinfo{year}{1996}), \eprint{hep-th/9606139}.

\bibitem[{\citenamefont{Ibanez et~al.}(2001)\citenamefont{Ibanez, Marchesano,
  and Rabadan}}]{Ibanez:2001nd}
\bibinfo{author}{\bibfnamefont{L.~E.} \bibnamefont{Ibanez}},
  \bibinfo{author}{\bibfnamefont{F.}~\bibnamefont{Marchesano}},
  \bibnamefont{and} \bibinfo{author}{\bibfnamefont{R.}~\bibnamefont{Rabadan}},
  {``}\bibinfo{title}{{Getting just the standard model at intersecting
  branes}},{''} \bibinfo{journal}{JHEP} \textbf{\bibinfo{volume}{0111}},
  \bibinfo{pages}{002} (\bibinfo{year}{2001}), \eprint{hep-th/0105155}.

\bibitem[{\citenamefont{Blumenhagen et~al.}(2001)\citenamefont{Blumenhagen,
  Kors, Lust, and Ott}}]{Blumenhagen:2001te}
\bibinfo{author}{\bibfnamefont{R.}~\bibnamefont{Blumenhagen}},
  \bibinfo{author}{\bibfnamefont{B.}~\bibnamefont{Kors}},
  \bibinfo{author}{\bibfnamefont{D.}~\bibnamefont{Lust}}, \bibnamefont{and}
  \bibinfo{author}{\bibfnamefont{T.}~\bibnamefont{Ott}},
  {``}\bibinfo{title}{{The standard model from stable intersecting brane world
  orbifolds}},{''} \bibinfo{journal}{Nucl.Phys.}
  \textbf{\bibinfo{volume}{B616}}, \bibinfo{pages}{3} (\bibinfo{year}{2001}),
  \eprint{hep-th/0107138}.

\bibitem[{\citenamefont{Cvetic et~al.}(2001{\natexlab{a}})\citenamefont{Cvetic,
  Shiu, and Uranga}}]{Cvetic:2001tj}
\bibinfo{author}{\bibfnamefont{M.}~\bibnamefont{Cvetic}},
  \bibinfo{author}{\bibfnamefont{G.}~\bibnamefont{Shiu}}, \bibnamefont{and}
  \bibinfo{author}{\bibfnamefont{A.~M.} \bibnamefont{Uranga}},
  {``}\bibinfo{title}{{Three family supersymmetric standard - like models from
  intersecting brane worlds}},{''} \bibinfo{journal}{Phys.Rev.Lett.}
  \textbf{\bibinfo{volume}{87}}, \bibinfo{pages}{201801}
  (\bibinfo{year}{2001}{\natexlab{a}}), \eprint{hep-th/0107143}.

\bibitem[{\citenamefont{Cvetic et~al.}(2001{\natexlab{b}})\citenamefont{Cvetic,
  Shiu, and Uranga}}]{Cvetic:2001nr}
\bibinfo{author}{\bibfnamefont{M.}~\bibnamefont{Cvetic}},
  \bibinfo{author}{\bibfnamefont{G.}~\bibnamefont{Shiu}}, \bibnamefont{and}
  \bibinfo{author}{\bibfnamefont{A.~M.} \bibnamefont{Uranga}},
  {``}\bibinfo{title}{{Chiral four-dimensional N=1 supersymmetric type 2A
  orientifolds from intersecting D6 branes}},{''} \bibinfo{journal}{Nucl.Phys.}
  \textbf{\bibinfo{volume}{B615}}, \bibinfo{pages}{3}
  (\bibinfo{year}{2001}{\natexlab{b}}), \eprint{hep-th/0107166}.

\bibitem[{\citenamefont{Cvetic et~al.}(2003)\citenamefont{Cvetic,
  Papadimitriou, and Shiu}}]{Cvetic:2002pj}
\bibinfo{author}{\bibfnamefont{M.}~\bibnamefont{Cvetic}},
  \bibinfo{author}{\bibfnamefont{I.}~\bibnamefont{Papadimitriou}},
  \bibnamefont{and} \bibinfo{author}{\bibfnamefont{G.}~\bibnamefont{Shiu}},
  {``}\bibinfo{title}{{Supersymmetric three family SU(5) grand unified models
  from type IIA orientifolds with intersecting D6-branes}},{''}
  \bibinfo{journal}{Nucl.Phys.} \textbf{\bibinfo{volume}{B659}},
  \bibinfo{pages}{193} (\bibinfo{year}{2003}), \eprint{hep-th/0212177}.

\bibitem[{\citenamefont{Cvetic et~al.}(2004)\citenamefont{Cvetic, Li, and
  Liu}}]{Cvetic:2004ui}
\bibinfo{author}{\bibfnamefont{M.}~\bibnamefont{Cvetic}},
  \bibinfo{author}{\bibfnamefont{T.}~\bibnamefont{Li}}, \bibnamefont{and}
  \bibinfo{author}{\bibfnamefont{T.}~\bibnamefont{Liu}},
  {``}\bibinfo{title}{{Supersymmetric patiSalam models from intersecting
  D6-branes: A Road to the standard model}},{''} \bibinfo{journal}{Nucl.Phys.}
  \textbf{\bibinfo{volume}{B698}}, \bibinfo{pages}{163} (\bibinfo{year}{2004}),
  \eprint{hep-th/0403061}.

\bibitem[{\citenamefont{Cvetic et~al.}(2005)\citenamefont{Cvetic, Langacker,
  Li, and Liu}}]{Cvetic:2004nk}
\bibinfo{author}{\bibfnamefont{M.}~\bibnamefont{Cvetic}},
  \bibinfo{author}{\bibfnamefont{P.}~\bibnamefont{Langacker}},
  \bibinfo{author}{\bibfnamefont{T.-j.} \bibnamefont{Li}}, \bibnamefont{and}
  \bibinfo{author}{\bibfnamefont{T.}~\bibnamefont{Liu}},
  {``}\bibinfo{title}{{D6-brane splitting on type IIA orientifolds}},{''}
  \bibinfo{journal}{Nucl.Phys.} \textbf{\bibinfo{volume}{B709}},
  \bibinfo{pages}{241} (\bibinfo{year}{2005}), \eprint{hep-th/0407178}.

\bibitem[{\citenamefont{Chen et~al.}(2005{\natexlab{a}})\citenamefont{Chen,
  Kraniotis, Mayes, Nanopoulos, and Walker}}]{Chen:2005aba}
\bibinfo{author}{\bibfnamefont{C.-M.} \bibnamefont{Chen}},
  \bibinfo{author}{\bibfnamefont{G.}~\bibnamefont{Kraniotis}},
  \bibinfo{author}{\bibfnamefont{V.}~\bibnamefont{Mayes}},
  \bibinfo{author}{\bibfnamefont{D.~V.} \bibnamefont{Nanopoulos}},
  \bibnamefont{and} \bibinfo{author}{\bibfnamefont{J.}~\bibnamefont{Walker}},
  {``}\bibinfo{title}{{A Supersymmetric flipped SU(5) intersecting brane
  world}},{''} \bibinfo{journal}{Phys.Lett.} \textbf{\bibinfo{volume}{B611}},
  \bibinfo{pages}{156} (\bibinfo{year}{2005}{\natexlab{a}}),
  \eprint{hep-th/0501182}.

\bibitem[{\citenamefont{Chen et~al.}(2005{\natexlab{b}})\citenamefont{Chen,
  Kraniotis, Mayes, Nanopoulos, and Walker}}]{Chen:2005mm}
\bibinfo{author}{\bibfnamefont{C.-M.} \bibnamefont{Chen}},
  \bibinfo{author}{\bibfnamefont{G.}~\bibnamefont{Kraniotis}},
  \bibinfo{author}{\bibfnamefont{V.}~\bibnamefont{Mayes}},
  \bibinfo{author}{\bibfnamefont{D.~V.} \bibnamefont{Nanopoulos}},
  \bibnamefont{and} \bibinfo{author}{\bibfnamefont{J.}~\bibnamefont{Walker}},
  {``}\bibinfo{title}{{A K-theory anomaly free supersymmetric flipped SU(5)
  model from intersecting branes}},{''} \bibinfo{journal}{Phys.Lett.}
  \textbf{\bibinfo{volume}{B625}}, \bibinfo{pages}{96}
  (\bibinfo{year}{2005}{\natexlab{b}}), \eprint{hep-th/0507232}.

\bibitem[{\citenamefont{Chen et~al.}(2006)\citenamefont{Chen, Li, and
  Nanopoulos}}]{Chen:2005mj}
\bibinfo{author}{\bibfnamefont{C.-M.} \bibnamefont{Chen}},
  \bibinfo{author}{\bibfnamefont{T.}~\bibnamefont{Li}}, \bibnamefont{and}
  \bibinfo{author}{\bibfnamefont{D.~V.} \bibnamefont{Nanopoulos}},
  {``}\bibinfo{title}{{Standard-like model building on Type II
  orientifolds}},{''} \bibinfo{journal}{Nucl.Phys.}
  \textbf{\bibinfo{volume}{B732}}, \bibinfo{pages}{224} (\bibinfo{year}{2006}),
  \eprint{hep-th/0509059}.

\bibitem[{\citenamefont{Blumenhagen et~al.}(2005)\citenamefont{Blumenhagen,
  Cvetic, Langacker, and Shiu}}]{Blumenhagen:2005mu}
\bibinfo{author}{\bibfnamefont{R.}~\bibnamefont{Blumenhagen}},
  \bibinfo{author}{\bibfnamefont{M.}~\bibnamefont{Cvetic}},
  \bibinfo{author}{\bibfnamefont{P.}~\bibnamefont{Langacker}},
  \bibnamefont{and} \bibinfo{author}{\bibfnamefont{G.}~\bibnamefont{Shiu}},
  {``}\bibinfo{title}{{Toward realistic intersecting D-brane models}},{''}
  \bibinfo{journal}{Ann.Rev.Nucl.Part.Sci.} \textbf{\bibinfo{volume}{55}},
  \bibinfo{pages}{71} (\bibinfo{year}{2005}), \eprint{hep-th/0502005}.

\bibitem[{\citenamefont{Dijkstra
  et~al.}(2005{\natexlab{a}})\citenamefont{Dijkstra, Huiszoon, and
  Schellekens}}]{Dijkstra:2004ym}
\bibinfo{author}{\bibfnamefont{T.}~\bibnamefont{Dijkstra}},
  \bibinfo{author}{\bibfnamefont{L.}~\bibnamefont{Huiszoon}}, \bibnamefont{and}
  \bibinfo{author}{\bibfnamefont{A.}~\bibnamefont{Schellekens}},
  {``}\bibinfo{title}{{Chiral supersymmetric standard model spectra from
  orientifolds of Gepner models}},{''} \bibinfo{journal}{Phys.Lett.}
  \textbf{\bibinfo{volume}{B609}}, \bibinfo{pages}{408}
  (\bibinfo{year}{2005}{\natexlab{a}}), \eprint{hep-th/0403196}.

\bibitem[{\citenamefont{Dijkstra
  et~al.}(2005{\natexlab{b}})\citenamefont{Dijkstra, Huiszoon, and
  Schellekens}}]{Dijkstra:2004cc}
\bibinfo{author}{\bibfnamefont{T.}~\bibnamefont{Dijkstra}},
  \bibinfo{author}{\bibfnamefont{L.}~\bibnamefont{Huiszoon}}, \bibnamefont{and}
  \bibinfo{author}{\bibfnamefont{A.}~\bibnamefont{Schellekens}},
  {``}\bibinfo{title}{{Supersymmetric standard model spectra from RCFT
  orientifolds}},{''} \bibinfo{journal}{Nucl.Phys.}
  \textbf{\bibinfo{volume}{B710}}, \bibinfo{pages}{3}
  (\bibinfo{year}{2005}{\natexlab{b}}), \eprint{hep-th/0411129}.

\bibitem[{\citenamefont{Acharya and Witten}(2001)}]{Acharya:2001gy}
\bibinfo{author}{\bibfnamefont{B.~S.} \bibnamefont{Acharya}} \bibnamefont{and}
  \bibinfo{author}{\bibfnamefont{E.}~\bibnamefont{Witten}},
  {``}\bibinfo{title}{{Chiral fermions from manifolds of G(2) holonomy}},{''}
  (\bibinfo{year}{2001}), \eprint{hep-th/0109152}.

\bibitem[{\citenamefont{Friedmann and Witten}(2003)}]{Friedmann:2002ty}
\bibinfo{author}{\bibfnamefont{T.}~\bibnamefont{Friedmann}} \bibnamefont{and}
  \bibinfo{author}{\bibfnamefont{E.}~\bibnamefont{Witten}},
  {``}\bibinfo{title}{{Unification scale, proton decay, and manifolds of G(2)
  holonomy}},{''} \bibinfo{journal}{Adv.Theor.Math.Phys.}
  \textbf{\bibinfo{volume}{7}}, \bibinfo{pages}{577} (\bibinfo{year}{2003}),
  \eprint{hep-th/0211269}.

\bibitem[{\citenamefont{Beasley
  et~al.}(2009{\natexlab{a}})\citenamefont{Beasley, Heckman, and
  Vafa}}]{Beasley:2008dc}
\bibinfo{author}{\bibfnamefont{C.}~\bibnamefont{Beasley}},
  \bibinfo{author}{\bibfnamefont{J.~J.} \bibnamefont{Heckman}},
  \bibnamefont{and} \bibinfo{author}{\bibfnamefont{C.}~\bibnamefont{Vafa}},
  {``}\bibinfo{title}{{GUTs and Exceptional Branes in F-theory - I}},{''}
  \bibinfo{journal}{JHEP} \textbf{\bibinfo{volume}{01}}, \bibinfo{pages}{058}
  (\bibinfo{year}{2009}{\natexlab{a}}), \eprint{0802.3391}.

\bibitem[{\citenamefont{Beasley
  et~al.}(2009{\natexlab{b}})\citenamefont{Beasley, Heckman, and
  Vafa}}]{Beasley:2008kw}
\bibinfo{author}{\bibfnamefont{C.}~\bibnamefont{Beasley}},
  \bibinfo{author}{\bibfnamefont{J.~J.} \bibnamefont{Heckman}},
  \bibnamefont{and} \bibinfo{author}{\bibfnamefont{C.}~\bibnamefont{Vafa}},
  {``}\bibinfo{title}{{GUTs and Exceptional Branes in F-theory - II:
  Experimental Predictions}},{''} \bibinfo{journal}{JHEP}
  \textbf{\bibinfo{volume}{01}}, \bibinfo{pages}{059}
  (\bibinfo{year}{2009}{\natexlab{b}}), \eprint{0806.0102}.

\bibitem[{\citenamefont{Donagi and
  Wijnholt}(2008{\natexlab{a}})}]{Donagi:2008ca}
\bibinfo{author}{\bibfnamefont{R.}~\bibnamefont{Donagi}} \bibnamefont{and}
  \bibinfo{author}{\bibfnamefont{M.}~\bibnamefont{Wijnholt}},
  {``}\bibinfo{title}{{Model Building with F-Theory}},{''}
  (\bibinfo{year}{2008}{\natexlab{a}}), \eprint{0802.2969}.

\bibitem[{\citenamefont{Donagi and
  Wijnholt}(2008{\natexlab{b}})}]{Donagi:2008kj}
\bibinfo{author}{\bibfnamefont{R.}~\bibnamefont{Donagi}} \bibnamefont{and}
  \bibinfo{author}{\bibfnamefont{M.}~\bibnamefont{Wijnholt}},
  {``}\bibinfo{title}{{Breaking GUT Groups in F-Theory}},{''}
  (\bibinfo{year}{2008}{\natexlab{b}}), \eprint{0808.2223}.

\bibitem[{\citenamefont{Jiang et~al.}(2009)\citenamefont{Jiang, Li, Nanopoulos,
  and Xie}}]{Jiang:2009zza}
\bibinfo{author}{\bibfnamefont{J.}~\bibnamefont{Jiang}},
  \bibinfo{author}{\bibfnamefont{T.}~\bibnamefont{Li}},
  \bibinfo{author}{\bibfnamefont{D.~V.} \bibnamefont{Nanopoulos}},
  \bibnamefont{and} \bibinfo{author}{\bibfnamefont{D.}~\bibnamefont{Xie}},
  {``}\bibinfo{title}{{F-$SU(5)$}},{''} \bibinfo{journal}{Phys. Lett.}
  \textbf{\bibinfo{volume}{B677}}, \bibinfo{pages}{322} (\bibinfo{year}{2009}).

\bibitem[{\citenamefont{Jiang et~al.}(2010)\citenamefont{Jiang, Li, Nanopoulos,
  and Xie}}]{Jiang:2009za}
\bibinfo{author}{\bibfnamefont{J.}~\bibnamefont{Jiang}},
  \bibinfo{author}{\bibfnamefont{T.}~\bibnamefont{Li}},
  \bibinfo{author}{\bibfnamefont{D.~V.} \bibnamefont{Nanopoulos}},
  \bibnamefont{and} \bibinfo{author}{\bibfnamefont{D.}~\bibnamefont{Xie}},
  {``}\bibinfo{title}{{Flipped $SU(5) \times U(1)_X$ Models from
  F-Theory}},{''} \bibinfo{journal}{Nucl. Phys.}
  \textbf{\bibinfo{volume}{B830}}, \bibinfo{pages}{195} (\bibinfo{year}{2010}),
  \eprint{0905.3394}.

\bibitem[{\citenamefont{Bousso and Polchinski}(2000)}]{Bousso:2000xa}
\bibinfo{author}{\bibfnamefont{R.}~\bibnamefont{Bousso}} \bibnamefont{and}
  \bibinfo{author}{\bibfnamefont{J.}~\bibnamefont{Polchinski}},
  {``}\bibinfo{title}{{Quantization of four-form fluxes and dynamical
  neutralization of the cosmological constant}},{''} \bibinfo{journal}{JHEP}
  \textbf{\bibinfo{volume}{06}}, \bibinfo{pages}{006} (\bibinfo{year}{2000}),
  \eprint{hep-th/0004134}.

\bibitem[{\citenamefont{Giddings et~al.}(2002)\citenamefont{Giddings, Kachru,
  and Polchinski}}]{Giddings:2001yu}
\bibinfo{author}{\bibfnamefont{S.~B.} \bibnamefont{Giddings}},
  \bibinfo{author}{\bibfnamefont{S.}~\bibnamefont{Kachru}}, \bibnamefont{and}
  \bibinfo{author}{\bibfnamefont{J.}~\bibnamefont{Polchinski}},
  {``}\bibinfo{title}{{Hierarchies from fluxes in string
  compactifications}},{''} \bibinfo{journal}{Phys. Rev.}
  \textbf{\bibinfo{volume}{D66}}, \bibinfo{pages}{106006}
  (\bibinfo{year}{2002}), \eprint{hep-th/0105097}.

\bibitem[{\citenamefont{Kachru et~al.}(2003)\citenamefont{Kachru, Kallosh,
  Linde, and Trivedi}}]{Kachru:2003aw}
\bibinfo{author}{\bibfnamefont{S.}~\bibnamefont{Kachru}},
  \bibinfo{author}{\bibfnamefont{R.}~\bibnamefont{Kallosh}},
  \bibinfo{author}{\bibfnamefont{A.~D.} \bibnamefont{Linde}}, \bibnamefont{and}
  \bibinfo{author}{\bibfnamefont{S.~P.} \bibnamefont{Trivedi}},
  {``}\bibinfo{title}{{De Sitter vacua in string theory}},{''}
  \bibinfo{journal}{Phys. Rev.} \textbf{\bibinfo{volume}{D68}},
  \bibinfo{pages}{046005} (\bibinfo{year}{2003}), \eprint{hep-th/0301240}.

\bibitem[{\citenamefont{Susskind}(2003)}]{Susskind:2003kw}
\bibinfo{author}{\bibfnamefont{L.}~\bibnamefont{Susskind}},
  {``}\bibinfo{title}{{The Anthropic landscape of string theory}},{''}
  (\bibinfo{year}{2003}), \eprint{hep-th/0302219}.

\bibitem[{\citenamefont{Denef and Douglas}(2004)}]{Denef:2004ze}
\bibinfo{author}{\bibfnamefont{F.}~\bibnamefont{Denef}} \bibnamefont{and}
  \bibinfo{author}{\bibfnamefont{M.~R.} \bibnamefont{Douglas}},
  {``}\bibinfo{title}{{Distributions of flux vacua}},{''}
  \bibinfo{journal}{JHEP} \textbf{\bibinfo{volume}{0405}}, \bibinfo{pages}{072}
  (\bibinfo{year}{2004}), \eprint{hep-th/0404116}.

\bibitem[{\citenamefont{Denef and Douglas}(2005)}]{Denef:2004cf}
\bibinfo{author}{\bibfnamefont{F.}~\bibnamefont{Denef}} \bibnamefont{and}
  \bibinfo{author}{\bibfnamefont{M.~R.} \bibnamefont{Douglas}},
  {``}\bibinfo{title}{{Distributions of nonsupersymmetric flux vacua}},{''}
  \bibinfo{journal}{JHEP} \textbf{\bibinfo{volume}{0503}}, \bibinfo{pages}{061}
  (\bibinfo{year}{2005}), \eprint{hep-th/0411183}.

\bibitem[{\citenamefont{Barr}(1982)}]{Barr:1981qv}
\bibinfo{author}{\bibfnamefont{S.~M.} \bibnamefont{Barr}},
  {``}\bibinfo{title}{{A New Symmetry Breaking Pattern for $SO(10)$ and Proton
  Decay}},{''} \bibinfo{journal}{Phys. Lett.} \textbf{\bibinfo{volume}{B112}},
  \bibinfo{pages}{219} (\bibinfo{year}{1982}).

\bibitem[{\citenamefont{Derendinger et~al.}(1984)\citenamefont{Derendinger,
  Kim, and Nanopoulos}}]{Derendinger:1983aj}
\bibinfo{author}{\bibfnamefont{J.~P.} \bibnamefont{Derendinger}},
  \bibinfo{author}{\bibfnamefont{J.~E.} \bibnamefont{Kim}}, \bibnamefont{and}
  \bibinfo{author}{\bibfnamefont{D.~V.} \bibnamefont{Nanopoulos}},
  {``}\bibinfo{title}{{Anti-$SU(5)$}},{''} \bibinfo{journal}{Phys. Lett.}
  \textbf{\bibinfo{volume}{B139}}, \bibinfo{pages}{170} (\bibinfo{year}{1984}).

\bibitem[{\citenamefont{Antoniadis et~al.}(1987)\citenamefont{Antoniadis,
  Ellis, Hagelin, and Nanopoulos}}]{Antoniadis:1987dx}
\bibinfo{author}{\bibfnamefont{I.}~\bibnamefont{Antoniadis}},
  \bibinfo{author}{\bibfnamefont{J.~R.} \bibnamefont{Ellis}},
  \bibinfo{author}{\bibfnamefont{J.~S.} \bibnamefont{Hagelin}},
  \bibnamefont{and} \bibinfo{author}{\bibfnamefont{D.~V.}
  \bibnamefont{Nanopoulos}}, {``}\bibinfo{title}{{Supersymmetric Flipped
  $SU(5)$ Revitalized}},{''} \bibinfo{journal}{Phys. Lett.}
  \textbf{\bibinfo{volume}{B194}}, \bibinfo{pages}{231} (\bibinfo{year}{1987}).

\bibitem[{\citenamefont{Jiang et~al.}(2007)\citenamefont{Jiang, Li, and
  Nanopoulos}}]{Jiang:2006hf}
\bibinfo{author}{\bibfnamefont{J.}~\bibnamefont{Jiang}},
  \bibinfo{author}{\bibfnamefont{T.}~\bibnamefont{Li}}, \bibnamefont{and}
  \bibinfo{author}{\bibfnamefont{D.~V.} \bibnamefont{Nanopoulos}},
  {``}\bibinfo{title}{{Testable Flipped $SU(5) \times U(1)_X$ Models}},{''}
  \bibinfo{journal}{Nucl. Phys.} \textbf{\bibinfo{volume}{B772}},
  \bibinfo{pages}{49} (\bibinfo{year}{2007}), \eprint{hep-ph/0610054}.

\bibitem[{\citenamefont{Cremmer et~al.}(1983)\citenamefont{Cremmer, Ferrara,
  Kounnas, and Nanopoulos}}]{Cremmer:1983bf}
\bibinfo{author}{\bibfnamefont{E.}~\bibnamefont{Cremmer}},
  \bibinfo{author}{\bibfnamefont{S.}~\bibnamefont{Ferrara}},
  \bibinfo{author}{\bibfnamefont{C.}~\bibnamefont{Kounnas}}, \bibnamefont{and}
  \bibinfo{author}{\bibfnamefont{D.~V.} \bibnamefont{Nanopoulos}},
  {``}\bibinfo{title}{{Naturally Vanishing Cosmological Constant in $N=1$
  Supergravity}},{''} \bibinfo{journal}{Phys. Lett.}
  \textbf{\bibinfo{volume}{B133}}, \bibinfo{pages}{61} (\bibinfo{year}{1983}).

\bibitem[{\citenamefont{Ellis et~al.}(1984{\natexlab{a}})\citenamefont{Ellis,
  Lahanas, Nanopoulos, and Tamvakis}}]{Ellis:1983sf}
\bibinfo{author}{\bibfnamefont{J.~R.} \bibnamefont{Ellis}},
  \bibinfo{author}{\bibfnamefont{A.~B.} \bibnamefont{Lahanas}},
  \bibinfo{author}{\bibfnamefont{D.~V.} \bibnamefont{Nanopoulos}},
  \bibnamefont{and} \bibinfo{author}{\bibfnamefont{K.}~\bibnamefont{Tamvakis}},
  {``}\bibinfo{title}{{No-Scale Supersymmetric Standard Model}},{''}
  \bibinfo{journal}{Phys. Lett.} \textbf{\bibinfo{volume}{B134}},
  \bibinfo{pages}{429} (\bibinfo{year}{1984}{\natexlab{a}}).

\bibitem[{\citenamefont{Ellis et~al.}(1984{\natexlab{b}})\citenamefont{Ellis,
  Kounnas, and Nanopoulos}}]{Ellis:1983ei}
\bibinfo{author}{\bibfnamefont{J.~R.} \bibnamefont{Ellis}},
  \bibinfo{author}{\bibfnamefont{C.}~\bibnamefont{Kounnas}}, \bibnamefont{and}
  \bibinfo{author}{\bibfnamefont{D.~V.} \bibnamefont{Nanopoulos}},
  {``}\bibinfo{title}{{Phenomenological $SU(1,1)$ Supergravity}},{''}
  \bibinfo{journal}{Nucl. Phys.} \textbf{\bibinfo{volume}{B241}},
  \bibinfo{pages}{406} (\bibinfo{year}{1984}{\natexlab{b}}).

\bibitem[{\citenamefont{Ellis et~al.}(1984{\natexlab{c}})\citenamefont{Ellis,
  Kounnas, and Nanopoulos}}]{Ellis:1984bm}
\bibinfo{author}{\bibfnamefont{J.~R.} \bibnamefont{Ellis}},
  \bibinfo{author}{\bibfnamefont{C.}~\bibnamefont{Kounnas}}, \bibnamefont{and}
  \bibinfo{author}{\bibfnamefont{D.~V.} \bibnamefont{Nanopoulos}},
  {``}\bibinfo{title}{{No Scale Supersymmetric Guts}},{''}
  \bibinfo{journal}{Nucl. Phys.} \textbf{\bibinfo{volume}{B247}},
  \bibinfo{pages}{373} (\bibinfo{year}{1984}{\natexlab{c}}).

\bibitem[{\citenamefont{Lahanas and Nanopoulos}(1987)}]{Lahanas:1986uc}
\bibinfo{author}{\bibfnamefont{A.~B.} \bibnamefont{Lahanas}} \bibnamefont{and}
  \bibinfo{author}{\bibfnamefont{D.~V.} \bibnamefont{Nanopoulos}},
  {``}\bibinfo{title}{{The Road to No Scale Supergravity}},{''}
  \bibinfo{journal}{Phys. Rept.} \textbf{\bibinfo{volume}{145}},
  \bibinfo{pages}{1} (\bibinfo{year}{1987}).

\bibitem[{\citenamefont{Ferrara et~al.}(1994)\citenamefont{Ferrara, Kounnas,
  and Zwirner}}]{Ferrara:1994kg}
\bibinfo{author}{\bibfnamefont{S.}~\bibnamefont{Ferrara}},
  \bibinfo{author}{\bibfnamefont{C.}~\bibnamefont{Kounnas}}, \bibnamefont{and}
  \bibinfo{author}{\bibfnamefont{F.}~\bibnamefont{Zwirner}},
  {``}\bibinfo{title}{{Mass formulae and natural hierarchy in string effective
  supergravities}},{''} \bibinfo{journal}{Nucl. Phys.}
  \textbf{\bibinfo{volume}{B429}}, \bibinfo{pages}{589} (\bibinfo{year}{1994}),
  \eprint{hep-th/9405188}.

\bibitem[{\citenamefont{Witten}(1985)}]{Witten:1985xb}
\bibinfo{author}{\bibfnamefont{E.}~\bibnamefont{Witten}},
  {``}\bibinfo{title}{{Dimensional Reduction of Superstring Models}},{''}
  \bibinfo{journal}{Phys.Lett.} \textbf{\bibinfo{volume}{B155}},
  \bibinfo{pages}{151} (\bibinfo{year}{1985}).

\bibitem[{\citenamefont{Li et~al.}(1997)\citenamefont{Li, Lopez, and
  Nanopoulos}}]{Li:1997sk}
\bibinfo{author}{\bibfnamefont{T.-j.} \bibnamefont{Li}},
  \bibinfo{author}{\bibfnamefont{J.~L.} \bibnamefont{Lopez}}, \bibnamefont{and}
  \bibinfo{author}{\bibfnamefont{D.~V.} \bibnamefont{Nanopoulos}},
  {``}\bibinfo{title}{{Compactifications of M theory and their phenomenological
  consequences}},{''} \bibinfo{journal}{Phys.Rev.}
  \textbf{\bibinfo{volume}{D56}}, \bibinfo{pages}{2602} (\bibinfo{year}{1997}),
  \eprint{hep-ph/9704247}.

\bibitem[{\citenamefont{Cremmer and Julia}(1979)}]{Cremmer:1979up}
\bibinfo{author}{\bibfnamefont{E.}~\bibnamefont{Cremmer}} \bibnamefont{and}
  \bibinfo{author}{\bibfnamefont{B.}~\bibnamefont{Julia}},
  {``}\bibinfo{title}{{The SO(8) Supergravity}},{''}
  \bibinfo{journal}{Nucl.Phys.} \textbf{\bibinfo{volume}{B159}},
  \bibinfo{pages}{141} (\bibinfo{year}{1979}).

\bibitem[{\citenamefont{Li et~al.}(2011{\natexlab{d}})\citenamefont{Li, Maxin,
  Nanopoulos, and Walker}}]{Li:2010ws}
\bibinfo{author}{\bibfnamefont{T.}~\bibnamefont{Li}},
  \bibinfo{author}{\bibfnamefont{J.~A.} \bibnamefont{Maxin}},
  \bibinfo{author}{\bibfnamefont{D.~V.} \bibnamefont{Nanopoulos}},
  \bibnamefont{and} \bibinfo{author}{\bibfnamefont{J.~W.}
  \bibnamefont{Walker}}, {``}\bibinfo{title}{{The Golden Point of No-Scale and
  No-Parameter ${\cal F}$-$SU(5)$}},{''} \bibinfo{journal}{Phys. Rev.}
  \textbf{\bibinfo{volume}{D83}}, \bibinfo{pages}{056015}
  (\bibinfo{year}{2011}{\natexlab{d}}), \eprint{1007.5100}.

\bibitem[{\citenamefont{Li et~al.}(2011{\natexlab{e}})\citenamefont{Li, Maxin,
  Nanopoulos, and Walker}}]{Li:2010mi}
\bibinfo{author}{\bibfnamefont{T.}~\bibnamefont{Li}},
  \bibinfo{author}{\bibfnamefont{J.~A.} \bibnamefont{Maxin}},
  \bibinfo{author}{\bibfnamefont{D.~V.} \bibnamefont{Nanopoulos}},
  \bibnamefont{and} \bibinfo{author}{\bibfnamefont{J.~W.}
  \bibnamefont{Walker}}, {``}\bibinfo{title}{{The Golden Strip of Correlated
  Top Quark, Gaugino, and Vectorlike Mass In No-Scale, No-Parameter
  F-SU(5)}},{''} \bibinfo{journal}{Phys. Lett.}
  \textbf{\bibinfo{volume}{B699}}, \bibinfo{pages}{164}
  (\bibinfo{year}{2011}{\natexlab{e}}), \eprint{1009.2981}.

\bibitem[{\citenamefont{Li et~al.}(2011{\natexlab{f}})\citenamefont{Li, Maxin,
  Nanopoulos, and Walker}}]{Li:2011hr}
\bibinfo{author}{\bibfnamefont{T.}~\bibnamefont{Li}},
  \bibinfo{author}{\bibfnamefont{J.~A.} \bibnamefont{Maxin}},
  \bibinfo{author}{\bibfnamefont{D.~V.} \bibnamefont{Nanopoulos}},
  \bibnamefont{and} \bibinfo{author}{\bibfnamefont{J.~W.}
  \bibnamefont{Walker}}, {``}\bibinfo{title}{{Ultra High Jet Signals from
  Stringy No-Scale Supergravity}},{''} (\bibinfo{year}{2011}{\natexlab{f}}),
  \eprint{1103.2362}.

\bibitem[{\citenamefont{Li et~al.}(2011{\natexlab{g}})\citenamefont{Li, Maxin,
  Nanopoulos, and Walker}}]{Maxin:2011hy}
\bibinfo{author}{\bibfnamefont{T.}~\bibnamefont{Li}},
  \bibinfo{author}{\bibfnamefont{J.~A.} \bibnamefont{Maxin}},
  \bibinfo{author}{\bibfnamefont{D.~V.} \bibnamefont{Nanopoulos}},
  \bibnamefont{and} \bibinfo{author}{\bibfnamefont{J.~W.}
  \bibnamefont{Walker}}, {``}\bibinfo{title}{{The Ultra-High Jet Multiplicity
  Signal of Stringy No-Scale F-SU(5) at the $\sqrt{s}$ = 7 TeV LHC}},{''}
  \bibinfo{journal}{Phys. Rev.} \textbf{\bibinfo{volume}{D84}},
  \bibinfo{pages}{076003} (\bibinfo{year}{2011}{\natexlab{g}}),
  \eprint{1103.4160}.

\bibitem[{\citenamefont{Li et~al.}(2011{\natexlab{h}})\citenamefont{Li, Maxin,
  Nanopoulos, and Walker}}]{Li:2011in}
\bibinfo{author}{\bibfnamefont{T.}~\bibnamefont{Li}},
  \bibinfo{author}{\bibfnamefont{J.~A.} \bibnamefont{Maxin}},
  \bibinfo{author}{\bibfnamefont{D.~V.} \bibnamefont{Nanopoulos}},
  \bibnamefont{and} \bibinfo{author}{\bibfnamefont{J.~W.}
  \bibnamefont{Walker}}, {``}\bibinfo{title}{{The Race for Supersymmetric Dark
  Matter at XENON100 and the LHC: Stringy Correlations from No-Scale
  \cal{F}-SU(5)}},{''} (\bibinfo{year}{2011}{\natexlab{h}}),
  \eprint{1106.1165}.

\bibitem[{\citenamefont{Li et~al.}(2011{\natexlab{i}})\citenamefont{Li, Maxin,
  Nanopoulos, and Walker}}]{Li:2011gh}
\bibinfo{author}{\bibfnamefont{T.}~\bibnamefont{Li}},
  \bibinfo{author}{\bibfnamefont{J.~A.} \bibnamefont{Maxin}},
  \bibinfo{author}{\bibfnamefont{D.~V.} \bibnamefont{Nanopoulos}},
  \bibnamefont{and} \bibinfo{author}{\bibfnamefont{J.~W.}
  \bibnamefont{Walker}}, {``}\bibinfo{title}{{A Two-Tiered Correlation of Dark
  Matter with Missing Transverse Energy: Reconstructing the Lightest
  Supersymmetric Particle Mass at the LHC}},{''}
  (\bibinfo{year}{2011}{\natexlab{i}}), \eprint{1107.2375}.

\bibitem[{\citenamefont{Li et~al.}(2011{\natexlab{j}})\citenamefont{Li, Maxin,
  Nanopoulos, and Walker}}]{Li:2011rp}
\bibinfo{author}{\bibfnamefont{T.}~\bibnamefont{Li}},
  \bibinfo{author}{\bibfnamefont{J.~A.} \bibnamefont{Maxin}},
  \bibinfo{author}{\bibfnamefont{D.~V.} \bibnamefont{Nanopoulos}},
  \bibnamefont{and} \bibinfo{author}{\bibfnamefont{J.~W.}
  \bibnamefont{Walker}}, {``}\bibinfo{title}{{Prospects for Discovery of
  Supersymmetric No-Scale F-SU(5) at The Once and Future LHC}},{''}
  (\bibinfo{year}{2011}{\natexlab{j}}), \eprint{1107.3825}.

\bibitem[{\citenamefont{Li et~al.}(2011{\natexlab{k}})\citenamefont{Li, Maxin,
  Nanopoulos, and Walker}}]{Li:2011fu}
\bibinfo{author}{\bibfnamefont{T.}~\bibnamefont{Li}},
  \bibinfo{author}{\bibfnamefont{J.~A.} \bibnamefont{Maxin}},
  \bibinfo{author}{\bibfnamefont{D.~V.} \bibnamefont{Nanopoulos}},
  \bibnamefont{and} \bibinfo{author}{\bibfnamefont{J.~W.}
  \bibnamefont{Walker}}, {``}\bibinfo{title}{{Has SUSY Gone Undetected in 9-jet
  Events? A Ten-Fold Enhancement in the LHC Signal Efficiency}},{''}
  (\bibinfo{year}{2011}{\natexlab{k}}), \eprint{1108.5169}.

\bibitem[{\citenamefont{Li et~al.}(2011{\natexlab{l}})\citenamefont{Li, Maxin,
  Nanopoulos, and Walker}}]{Li:2011xg}
\bibinfo{author}{\bibfnamefont{T.}~\bibnamefont{Li}},
  \bibinfo{author}{\bibfnamefont{J.~A.} \bibnamefont{Maxin}},
  \bibinfo{author}{\bibfnamefont{D.~V.} \bibnamefont{Nanopoulos}},
  \bibnamefont{and} \bibinfo{author}{\bibfnamefont{J.~W.}
  \bibnamefont{Walker}}, {``}\bibinfo{title}{{A Natural Prediction for the
  Higgs Boson Mass: 120+3.5-1 GeV}},{''} (\bibinfo{year}{2011}{\natexlab{l}}),
  \eprint{1109.2110}.

\bibitem[{\citenamefont{Li et~al.}(2011{\natexlab{m}})\citenamefont{Li,
  Nanopoulos, and Walker}}]{Li:2010dp}
\bibinfo{author}{\bibfnamefont{T.}~\bibnamefont{Li}},
  \bibinfo{author}{\bibfnamefont{D.~V.} \bibnamefont{Nanopoulos}},
  \bibnamefont{and} \bibinfo{author}{\bibfnamefont{J.~W.}
  \bibnamefont{Walker}}, {``}\bibinfo{title}{{Elements of F-ast Proton
  Decay}},{''} \bibinfo{journal}{Nucl. Phys.} \textbf{\bibinfo{volume}{B846}},
  \bibinfo{pages}{43} (\bibinfo{year}{2011}{\natexlab{m}}), \eprint{1003.2570}.

\bibitem[{\citenamefont{Li et~al.}(2011{\natexlab{n}})\citenamefont{Li, Maxin,
  Nanopoulos, and Walker}}]{Li:2010rz}
\bibinfo{author}{\bibfnamefont{T.}~\bibnamefont{Li}},
  \bibinfo{author}{\bibfnamefont{J.~A.} \bibnamefont{Maxin}},
  \bibinfo{author}{\bibfnamefont{D.~V.} \bibnamefont{Nanopoulos}},
  \bibnamefont{and} \bibinfo{author}{\bibfnamefont{J.~W.}
  \bibnamefont{Walker}}, {``}\bibinfo{title}{{Dark Matter, Proton Decay and
  Other Phenomenological Constraints in ${\cal F}$-SU(5)}},{''}
  \bibinfo{journal}{Nucl.Phys.} \textbf{\bibinfo{volume}{B848}},
  \bibinfo{pages}{314} (\bibinfo{year}{2011}{\natexlab{n}}),
  \eprint{1003.4186}.

\bibitem[{\citenamefont{Li et~al.}(2010)\citenamefont{Li, Nanopoulos, and
  Walker}}]{Li:2009fq}
\bibinfo{author}{\bibfnamefont{T.}~\bibnamefont{Li}},
  \bibinfo{author}{\bibfnamefont{D.~V.} \bibnamefont{Nanopoulos}},
  \bibnamefont{and} \bibinfo{author}{\bibfnamefont{J.~W.}
  \bibnamefont{Walker}}, {``}\bibinfo{title}{{Fast Proton Decay}},{''}
  \bibinfo{journal}{Phys. Lett.} \textbf{\bibinfo{volume}{B693}},
  \bibinfo{pages}{580} (\bibinfo{year}{2010}), \eprint{0910.0860}.

\bibitem[{\citenamefont{Ellis et~al.}(1983{\natexlab{a}})\citenamefont{Ellis,
  Nanopoulos, and Tamvakis}}]{Ellis:1982wr}
\bibinfo{author}{\bibfnamefont{J.~R.} \bibnamefont{Ellis}},
  \bibinfo{author}{\bibfnamefont{D.~V.} \bibnamefont{Nanopoulos}},
  \bibnamefont{and} \bibinfo{author}{\bibfnamefont{K.}~\bibnamefont{Tamvakis}},
  {``}\bibinfo{title}{{Grand Unification in Simple Supergravity}},{''}
  \bibinfo{journal}{Phys. Lett.} \textbf{\bibinfo{volume}{B121}},
  \bibinfo{pages}{123} (\bibinfo{year}{1983}{\natexlab{a}}).

\bibitem[{\citenamefont{Ellis et~al.}(1983{\natexlab{b}})\citenamefont{Ellis,
  Hagelin, Nanopoulos, and Tamvakis}}]{Ellis:1983bp}
\bibinfo{author}{\bibfnamefont{J.~R.} \bibnamefont{Ellis}},
  \bibinfo{author}{\bibfnamefont{J.}~\bibnamefont{Hagelin}},
  \bibinfo{author}{\bibfnamefont{D.~V.} \bibnamefont{Nanopoulos}},
  \bibnamefont{and} \bibinfo{author}{\bibfnamefont{K.}~\bibnamefont{Tamvakis}},
  {``}\bibinfo{title}{{Weak Symmetry Breaking by Radiative Corrections in
  Broken Supergravity}},{''} \bibinfo{journal}{Phys. Lett.}
  \textbf{\bibinfo{volume}{B125}}, \bibinfo{pages}{275}
  (\bibinfo{year}{1983}{\natexlab{b}}).

\bibitem[{\citenamefont{Djouadi et~al.}(2007)\citenamefont{Djouadi, Kneur, and
  Moultaka}}]{Djouadi:2002ze}
\bibinfo{author}{\bibfnamefont{A.}~\bibnamefont{Djouadi}},
  \bibinfo{author}{\bibfnamefont{J.-L.} \bibnamefont{Kneur}}, \bibnamefont{and}
  \bibinfo{author}{\bibfnamefont{G.}~\bibnamefont{Moultaka}},
  {``}\bibinfo{title}{{SuSpect: A Fortran code for the supersymmetric and Higgs
  particle spectrum in the MSSM}},{''} \bibinfo{journal}{Comput. Phys. Commun.}
  \textbf{\bibinfo{volume}{176}}, \bibinfo{pages}{426} (\bibinfo{year}{2007}),
  \eprint{hep-ph/0211331}.

\bibitem[{:19(2010)}]{:1900yx}
{``}\bibinfo{title}{{Combination of CDF and D0 Results on the Mass of the Top
  Quark using up to 5.6 $fb^{-1}$ of data (The CDF and D0 Collaboration)}},{''}
  (\bibinfo{year}{2010}), \eprint{1007.3178}.

\bibitem[{\citenamefont{Komatsu et~al.}(2010)}]{Komatsu:2010fb}
\bibinfo{author}{\bibfnamefont{E.}~\bibnamefont{Komatsu}} \bibnamefont{et~al.}
  (\bibinfo{collaboration}{WMAP}), {``}\bibinfo{title}{{Seven-Year Wilkinson
  Microwave Anisotropy Probe (WMAP) Observations: Cosmological
  Interpretation}},{''} \bibinfo{journal}{Astrophys.J.Suppl.}
  \textbf{\bibinfo{volume}{192}}, \bibinfo{pages}{18} (\bibinfo{year}{2010}),
  \eprint{1001.4538}.

\bibitem[{\citenamefont{Barate et~al.}(2003)}]{Barate:2003sz}
\bibinfo{author}{\bibfnamefont{R.}~\bibnamefont{Barate}} \bibnamefont{et~al.}
  (\bibinfo{collaboration}{LEP Working Group for Higgs boson searches}),
  {``}\bibinfo{title}{{Search for the standard model Higgs boson at LEP}},{''}
  \bibinfo{journal}{Phys. Lett.} \textbf{\bibinfo{volume}{B565}},
  \bibinfo{pages}{61} (\bibinfo{year}{2003}), \eprint{hep-ex/0306033}.

\bibitem[{\citenamefont{Yao et~al.}(2006)}]{Yao:2006px}
\bibinfo{author}{\bibfnamefont{W.~M.} \bibnamefont{Yao}} \bibnamefont{et~al.}
  (\bibinfo{collaboration}{Particle Data Group}), {``}\bibinfo{title}{{Review
  of Particle physics}},{''} \bibinfo{journal}{J. Phys.}
  \textbf{\bibinfo{volume}{G33}}, \bibinfo{pages}{1} (\bibinfo{year}{2006}).

\bibitem[{\citenamefont{Li et~al.}(2011{\natexlab{o}})\citenamefont{Li, Maxin,
  Nanopoulos, and Walker}}]{LMNW-P}
\bibinfo{author}{\bibfnamefont{T.}~\bibnamefont{Li}},
  \bibinfo{author}{\bibfnamefont{J.~A.} \bibnamefont{Maxin}},
  \bibinfo{author}{\bibfnamefont{D.~V.} \bibnamefont{Nanopoulos}},
  \bibnamefont{and} \bibinfo{author}{\bibfnamefont{J.~W.} \bibnamefont{Walker}}
  (\bibinfo{year}{2011}{\natexlab{o}}), \bibinfo{note}{in Preparation}.

\end{thebibliography}

\end{document}